\documentclass[8pt]{article}
\usepackage[utf8]{inputenc}
\usepackage{geometry}
\usepackage{amsmath}
\usepackage{amsfonts}
\usepackage{mathrsfs}
\usepackage{amsthm}
\usepackage{amssymb}
\usepackage{hyperref}
\usepackage[capitalize]{cleveref}
\usepackage{extarrows}
\usepackage{enumitem}
\usepackage{bbm}
\usepackage{upgreek}
\usepackage{mathtools}
\usepackage{wasysym}
\usepackage{xfrac}
\usepackage{graphicx}
\usepackage{multirow}
\usepackage{algorithm}
\usepackage{algpseudocode}
\usepackage[tableposition = top]{caption}
\usepackage{subcaption}
\usepackage{authblk}

\theoremstyle{definition}

\theoremstyle{remark}

\newcommand{\N}{\mathbb{N}}
\newcommand{\R}{\mathbb{R}}

\newcommand{\veps}{\varepsilon}
\newcommand{\eps}{\epsilon}
\newcommand{\pa}{\partial}

\DeclareMathOperator{\Tr}{Tr}
\renewcommand{\d}[1]{\ensuremath{\operatorname{d}\!{#1}}}

\title{Solving Non-local Fokker-Planck Equations by Deep Learning}
\author[1]{Senbao Jiang\thanks{Corresponding Author: sjiang23@hawk.iit.edu}}
\author[1]{Xiaofan Li\thanks{lix@iit.edu}}
\affil[1]{Department of Applied Mathematics, Illinois Institute of Technology}

\begin{document}

\maketitle

\begin{abstract}
   Physics-informed neural networks (PiNNs) recently emerged as a powerful solver for a large class of partial differential equations under various inital and boundary conditions. In this paper, we propose \emph{trapz-PiNNs}, physics-informed neural networks incoporated with a modified trapezoidal rule recently developed for accurately evaluating fractional laplacian and solve the space-fractional Fokker-Planck equations in 2D and 3D. We describe the modified trapezoidal rule in detail and verify the second-order accuracy. We demonstrate trapz-PiNNs have high expressive power through predicting solution with low $\mathcal{L}^2$ relative error on a varitey of numerical examples. We also use local metrics such as pointwise absolute and relative errors to analyze where could be further improved. We present an effective method for improving performance of trapz-PiNN on local metrics, provided that physical observations of high-fidelity simulation of the true solution are available. Besides the usual advantages of the deep learning solvers such as adaptivity and mesh-independence, the trapz-PiNN is able to solve PDEs with fractional laplacian with arbitrary $\alpha\in (0,2)$ and specializes on rectangular domain. It also has potential to be generlized into higher dimensions.
\end{abstract}

%%%%%%%%%%%%%%%%%%%%%%%%%%%%%%%%%%%%%%%%%%%%%%%%%%%%%%%%%%%%%%%%%%%%%%%%%%%%%%%%%%%%%%%%%%%%%%%%%%%%%%%%%%%%%%%%%%%%%%%%%%%%%%%%%%%%%%%%%%%%%%%%%%%%%%%%%%%%%%%%%%%%%%%%%%%%%%%%%%%%%%%%%%%%%%%%%%%%%%%%%%%%%%%%%%%%%%%%%%%%%%%%%%%%%%%%%%%%%%%%%%%%%%%%%%%%%%%%%%%%%%%%%%%%%%%%%%%%%%%%%%%%%%%%%%%%%%%%%%%%%%%%%%%%%%%%%%%%
\section{Introduction}
The Fokker-Planck equations (FPEs) describe the time evolution of probability density functions of underlying stochastic dynamics \cite{duan2015introduction}. If the driving noise is Gaussian (brownian motions), the FPE is a parabolic partial differential equation involving laplacian. The FPEs are widely used in studying stochatic models in physical, chemical and biological systems. There are many cases that the driving noises are non-Gaussian \cite{2005AGUFMNG23B0097D,7353129,roberts2015heavy}, such as isotropic $\alpha$-stable L\'evy motions, then the corresponding FPEs are non-local parabolic PDE involving fractional laplacian. For a introduction to FPEs, see \cite{duan2015introduction}.

We are insterested in the following non-local Fokker-Planck Equation defined on $\Omega_T\coloneqq \Omega\times (0,T)$ where $\Omega\in\R^n,\ n\geq 1$ is a rectangular domain \begin{align}
    \partial_tu(x,t) &= -\nabla \cdot (f\ u) + \frac{1}{2}\Tr\left\{\nabla^2: (\sigma(x)\sigma(x)^Tu(x,t)) \right\} - (-\Delta)^{\frac{\alpha}{2}}u(x,t),\quad (x,t)\in \Omega_T, \label{eq:Fokker-Planck} \\
    u(x,t) &= 0,\quad x\not \in \Omega,\ t>0, \label{eq:bdry cond}\\
    u(x,0) &= u_0(x). \label{eq:init cond}
\end{align} Here, $\alpha\in(0,2)$, $f(x)$ is a $C^2$ vector field in $\R^n$ and $\sigma(x)$ is a $C^2$ $n\times n$ matrix-valued map, the fractional laplacian is defined by \begin{align}
    (-\Delta)^{\frac{\alpha}{2}}\varphi(x) &\coloneqq c_{n,\alpha}\text{P.V.}\int_{\R^n\setminus\{0\}}\frac{\varphi(x)-\varphi(y)}{|x-y|^{n+\alpha}}\d y, \label{eq:frac lap}\\
    \text{where}\quad c_{n,\alpha} &\coloneqq \frac{2^{\alpha}\Gamma(\frac{\alpha+n}{2})}{\pi^{\frac{n}{2}}|\Gamma(-\frac{\alpha}{2})|}.
\end{align} We sometimes denote \cref{eq:Fokker-Planck} by $ \pa_t u = \mathcal{L}u$ for convenience. After rescaling, we can always assume $\Omega = (-1,1)^n$, a hypercube with side length two. For more information about fractional laplacian, we refer to \cite{LISCHKE2020109009}.

Finite difference type numerical method for the non-local partial differential equations involving fractional laplacian usually relies on the discretization of fractional laplacian \cite{zhao2018fast, doi:10.1137/12086491X, wang2013fast,DUO2019639,GAO20161,HansenHa}. Recently, a new accurate numerical scheme has been developed for computing fractional laplacian in 2D \cite{JIANG2022127236}. It is a modified trapezoidal rule with correction terms around singularity in order to achieve high-order accuracy of the quadrature. The correction weights are pre-computed. Once being stored, the modified trapezoidal rule can be easily set up. Blended with finite difference method, it has been applied to solving one and two dimensional non-local FPEs in \cite{GAO20161,HansenHa} respectively. It is not hard to generalize this rule to arbitrary $n$ dimensions. For full presentation of the modified trapezoidal rule in 2D and its convergence analysis, please refer to \cite{JIANG2022127236}.

Apart from classical methods, the surge of artificial intelligence influenced numerical PDEs community. The Physics-informed Neural Networks (PiNNs) emerged as powerful deep learning solvers for partial differential equations (PDEs) \cite{SIRIGNANO20181339, doi:10.1073/pnas.1718942115, RAISSI2019686}, fractional PDEs \cite{PANG2020109760,doi:10.1137/18M1229845}, or stochastic PDEs \cite{doi:10.1137/19M1260141} with various initial and/or boundary conditions. PiNN was formally introduced by Rassi et al \cite{RAISSI2019686}, in which the authors provided data-driven solution (forward problem) and data-driven discovery (inverse problem) of some integer-order PDEs for continuous time models and discrete time models. PiNN has been further extended to solve nonlocal or fractional type PDEs by incorporating classcial numerical methods to evaluate non-local or fractional operators, where the automatic differentiation is not applicable. In \cite{PANG2020109760}, Pang et al proposed a unified nonlocal operator encompassing both fractional laplacian and classical laplacian, which is computed in spherical coordinates and Gauss-Legendre quadrature rule. Combining this unified operator with PiNN, they are able to solve non-local Poisson model and non-local turbulence. A parallel work by Pang et al \cite{doi:10.1137/18M1229845} combines PiNN with one-dimensional and multi-dimensional Grunwald-Letnikov numerical schemes and Gauss-Legendre quadrature rule to discretize the fractional laplacian and solves forward and inverse problem of fractional advection-diffusion equations in one, two and three dimensions in circular or spherical domains. Xu et al \cite{doi:10.1063/1.5132840} verifies the PiNN accurately solves the integer-order stationary FPEs in one, two and three dimenions.

In this work we propose the modified-trapezoidal-rule-incoporated Physics-informed Neural Networks and focus on forward problem for the non-local FPEs. We call \emph{trapz-PiNNs} for short, it is sufficient to only use the simplest version of the modified trapezoidal rule: a trapezoidal rule plus one correction term on singularity and provides second-order accuracy for evaluating the fractional laplacian. The expressive power of trapz-PiNN is demonstrated by accurately predicting the solution of nonlocal FPEs at time $T_{\text{pred}}$ after training strictly before the time $T$ for some $T\leq T_{\text{pred}}$. We verify the accuracy of the deep learning (DL) solution by comparing it with reference solution obtained from finite difference method (FDM) under $\mathcal{L}^2$ relative error. Apart from achieving good global metric $\mathcal{L}^2$ relative error, we investigate the DL solution profile from local metric such as pointwise absolute and relative errors. It turns out that even achieving good $\mathcal{L}^2$ relative error, trapz-PiNN predicts more accurately in the regions with large or moderate magnitude than regions with small magnitude. If physical observation of solution data or high-fidelity simulated data is available \cite{PANG2020109760}, we propse an effective loss function integrating the extra data into trapz-PiNN framework to furthur enhance the performance on the local metrics, provided physical observation of solution data or high-fidelity simulated data is available.

It worths mentioning that trapz-PiNN is not limited to being the solver of the non-local FPEs. It can be adapted to other non-local PDEs involving fractional laplacian such as fractional reaction-diffusion equations, Cahn-Hilliard equations etc. Besides the usual advantages of PiNNs such as independence of mesh grids and easy adaptivity, the proposed trapz-PiNN has the following characteristics: \begin{enumerate}
    \item trapz-PiNN solves the non-local PDEs on rectangular spacial domains while the previous relevent works emphasize on circular or spherical ones.
    \item The numerical scheme for fractional laplacian is valid for arbitrary $\alpha\in (0,2)$, and so is trapz-PiNN.
    \item Although we only provide numerical examples in two or three dimensions in this work, trapz-PiNN can be generalized to higher dimensional settings.
\end{enumerate}

We describe some notations that appear throughout the paper. $n$ denotes spatial dimensions, usually $2,3$ but could be higher. $\Delta t>0$ represents the time step size and $h>0$ means space resolution. $|x|$ is the $l^2$ norm of $x\in \R^n$. $\Omega = (-1,1)^n$ is a hypercube with side length two and $\Omega_T = \Omega\times (0,T)$ for some $T>0$. To facilitate reading, \cref{tab:acronyms} records all common acronyms used in this paper. 
%%%%%%%%%%%%%%%%%%%%%%%%%%%%%%%%%%%%%%%%%%%%%%%%%%%%%%%%%%%%%%%%%%%%%%%% acronyms table
\begin{table}[h]
    \centering
    \begin{tabular}{c|c}
    \hline\hline
    Full word & acronyms \\
    \hline
    Partial Differential Equations & PDEs \\
    Fokker-Planck Equations & FPEs \\
    Ornstein–Uhlenbeck & O-U \\
    Finite Difference Method & FDM\\
    Neural Network & NN \\
    Physics-informed Neural Networks & PiNNs \\
    Mean Squared Error & MSE \\
    Deep Learning & DL \\
    Stochastic Gradient Descent & SGD \\
     \hline\hline
    \end{tabular}
    \caption{ Common Acronyms}
    \label{tab:acronyms}
\end{table}
%%%%%%%%%%%%%%%%%%%%%%%%%%%%%%%%%%%%%%%%%%%%%%%%%%%%%%%%%%%%%%%%%%%%%%%%

We organize this paper as follows. In \cref{sec:num frac lap} we discuss the modified trapezoidal rule and demonstrate the second-order accuracy for approximating the fractional laplacian. We elaborate the construction of the trapz-PiNNs in \cref{sec:PiNNs}. Numerical results are given in detail in \cref{sec:num experiment}, in particular, we solve fractional heat equations in 2D and 3D and FPE with Ornstein–Uhlenbeck potential in 2D. We also proposed a modified loss function and compare the DL solution profiles computed by trapz-PiNNs equipped with the original and modified loss functions. \cref{sec:conclution} summarizes the work and present future directions.

%%%%%%%%%%%%%%%%%%%%%%%%%%%%%%%%%%%%%%%%%%%%%%%%%%%%%%%%%%%%%%%%%%%%%%%%%%%%%%%%%%%%%%%%%%%%%%%%%%%%%%%%%%%%%%%%%%%%%%%%%%%%%%%%%%%%%%%%%%%%%%%%%%%%%%%%%%%%%%%%%%%%%%%%%%%%%%%%%%%%%%%%%%%%%%%%%%%%%%%%%%%%%%%%%%%%%%%%%%%%%%%%%%%%%%%%%%%%%%%%%%%%%%%%%%%%%%%%%%%%%%%%%%%%%%%%%%%%%%%%%%%%%%%%%%%%%%%%%%%%%%%%%%%%%%%%

\section{Numerical Scheme for the Fractional Laplacian}\label{sec:num frac lap}
\subsection{The modified trapezoidal rule}
Accurately computing the fractional laplacian is essential in constructing the loss function in trapz-PiNNs. In this section, we introduce a simplest version of the modified trapezoidal rule introduced in \cite{JIANG2022127236}. Let $\phi\in C_c^N(\R^n)$ with $N>\max\{4-\alpha,n\}$. The modified trapezoidal rule is designed to numerically compute a class of weakly singular integrals, in particular \begin{align}
    I_{\alpha}^{(j)} = \int_{\R^n\setminus\{0\}}\phi(x)\frac{x_j^2}{|x|^{n+\alpha}}\d x,\quad j = 1,\cdots,n.\label{eq:singular integral}
\end{align} It turns out that discretization of fractional laplacian boils down to evaluating weakly singular integrals (\ref{eq:singular integral}). See \cite[section 3.3]{HansenHa}. Let $f$ be a Schwartz function or function with compact support on $\R^n$, we define punctured-hole trapezoidal rule to be \begin{align}
    T_h[f]\coloneqq h^n\sum_{k\in\mathbb{Z}^n,k\not = 0}f(kh).
\end{align} Writing \begin{align}
    s_j(x) = \frac{x_j^2}{|x|^{n+\alpha}}, \label{eq:singular_kernel}
\end{align} the modified trapezoidal rule for $I_{\alpha}^{(j)}$ is given by \begin{align}
    Q_h^{(j)}[\phi s_j] = T_h[\phi s_j] + h^{2-\alpha}\omega_0^{(j)} \phi(0). \label{eq:modified_trapz}
\end{align} Here, $\omega_0^{(j)}$ is a correction weight that can be computed by the following limit \begin{align}
    \omega_0^{(j)} = \lim_{h\to 0}\frac{1}{h^{2-\alpha}}\left( \int_{\R^n\setminus \{0\}} g(x)s_j(x)\d x - T_h[g \cdot s_j] \right),
\end{align} where $g$ is a radially symmetric Schwartz function. After a symmetry argument, one can see that \begin{align}
    \omega_0^{(1)} = \cdots = \omega_0^{(n)} \eqqcolon \omega_0.
\end{align} It is a proven fact that this simple modified trapezoidal rule has order of convergence $4-\alpha$ for weakly singular integral (\ref{eq:singular integral}). For more information about the full version of this numerical rule and related convergence analysis, please refer to \cite{JIANG2022127236}.

\subsection{Numerical Fractional Laplacian}
The modified trapezoidal rule (\ref{eq:modified_trapz}) can be applied to fractional laplacian (\ref{eq:frac lap}) with second-order accuracy. Let $\varphi(x)\in C_c^N(\R^n)$ with $supp(\varphi)\subset \Omega$ and let $\Omega_h \coloneqq (x_k)_k$ be any uniform mesh grid for $\Omega$ with mesh size $h$, then the fractional laplacian evaluated at $x_k$ is approximated by \begin{align}\label{eq:numerical_frac_lap}
    (-\Delta)^{\frac{\alpha}{2}}\varphi(x_k) &\approx h^n c_{n,\alpha}\sideset{}{''}\sum_{x_l\in\Omega_h\setminus\{x_k\}}\frac{\varphi(x_k)-\varphi(x_l)}{|x_k-x_l|^{n+\alpha}} - \frac{1}{2}c_{n,\alpha}\omega_0 h^{2-\alpha}\sum_{j=1}^n\partial_{jj}^2\varphi(x_k) \nonumber \\
    & + c_{n,\alpha}\varphi(x_k)I_{\alpha}(x_k),
\end{align} where $\sum_{x_l}^{''}$ indicates the summand is divided by $2^m$ if grid point $x_l$ lies on $m$ edges and \begin{align}
    I_{\alpha}(x) \coloneqq \int_{\Omega^c}\frac{\d y}{|y-x|^{n+\alpha}} \label{eq:analytical_part}
\end{align} can be evaluated analytically. In particular, the analytical formula for \cref{eq:analytical_part} in 2D is available in \cite{HansenHa}. We denote the discretized fractional laplacian (the RHS of \cref{eq:numerical_frac_lap}) by $(-\Delta)^{\sfrac{\alpha}{2}}_h$. 

\begin{figure}[h]
    \begin{subfigure}[b]{0.32\textwidth}
         \centering
         \includegraphics[width=\textwidth]{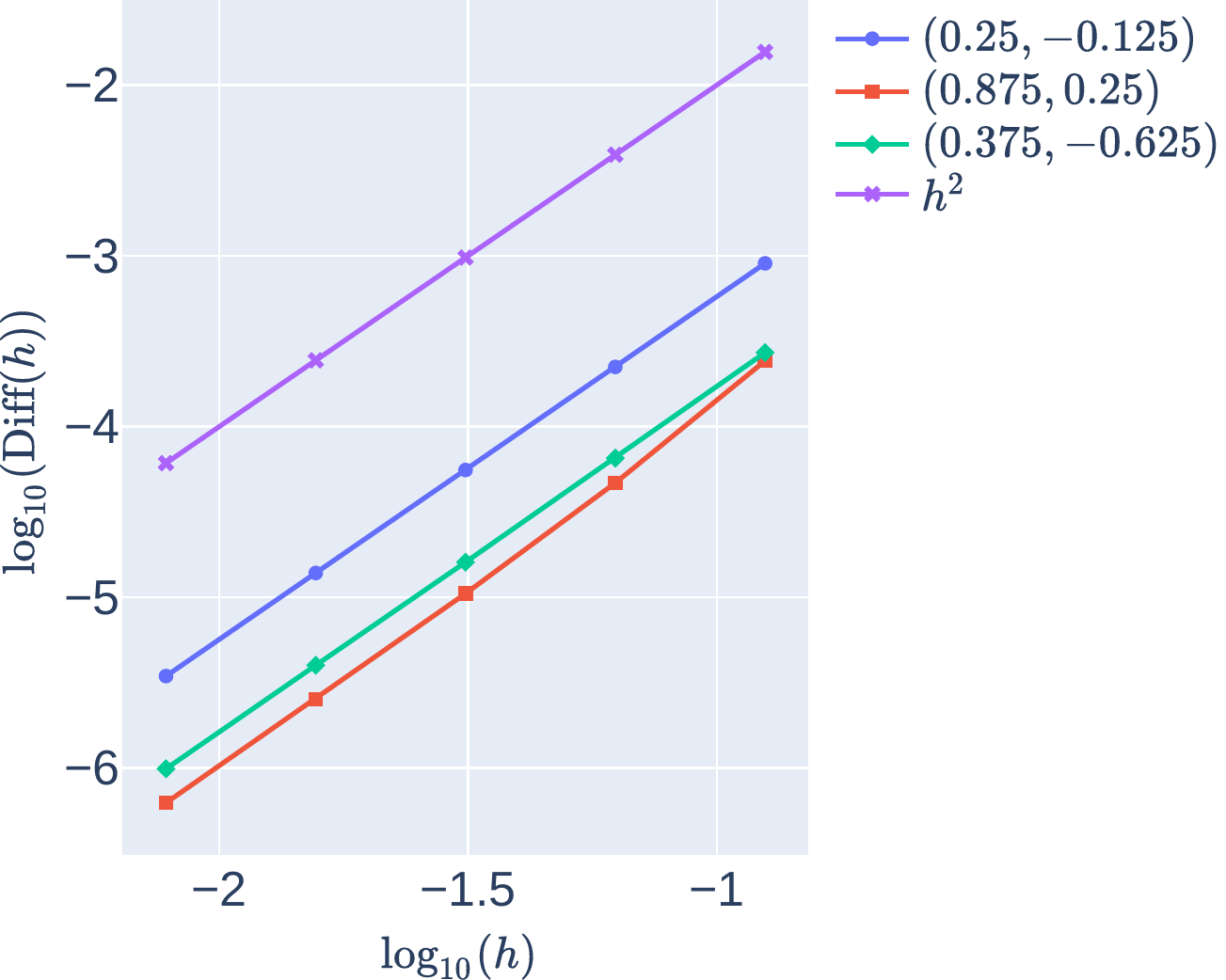}
         \caption{$\alpha = 0.5$ }
         \label{fig:alpha_half}
     \end{subfigure}
     \hfill
         \begin{subfigure}[b]{0.32\textwidth}
         \centering
         \includegraphics[width=\textwidth]{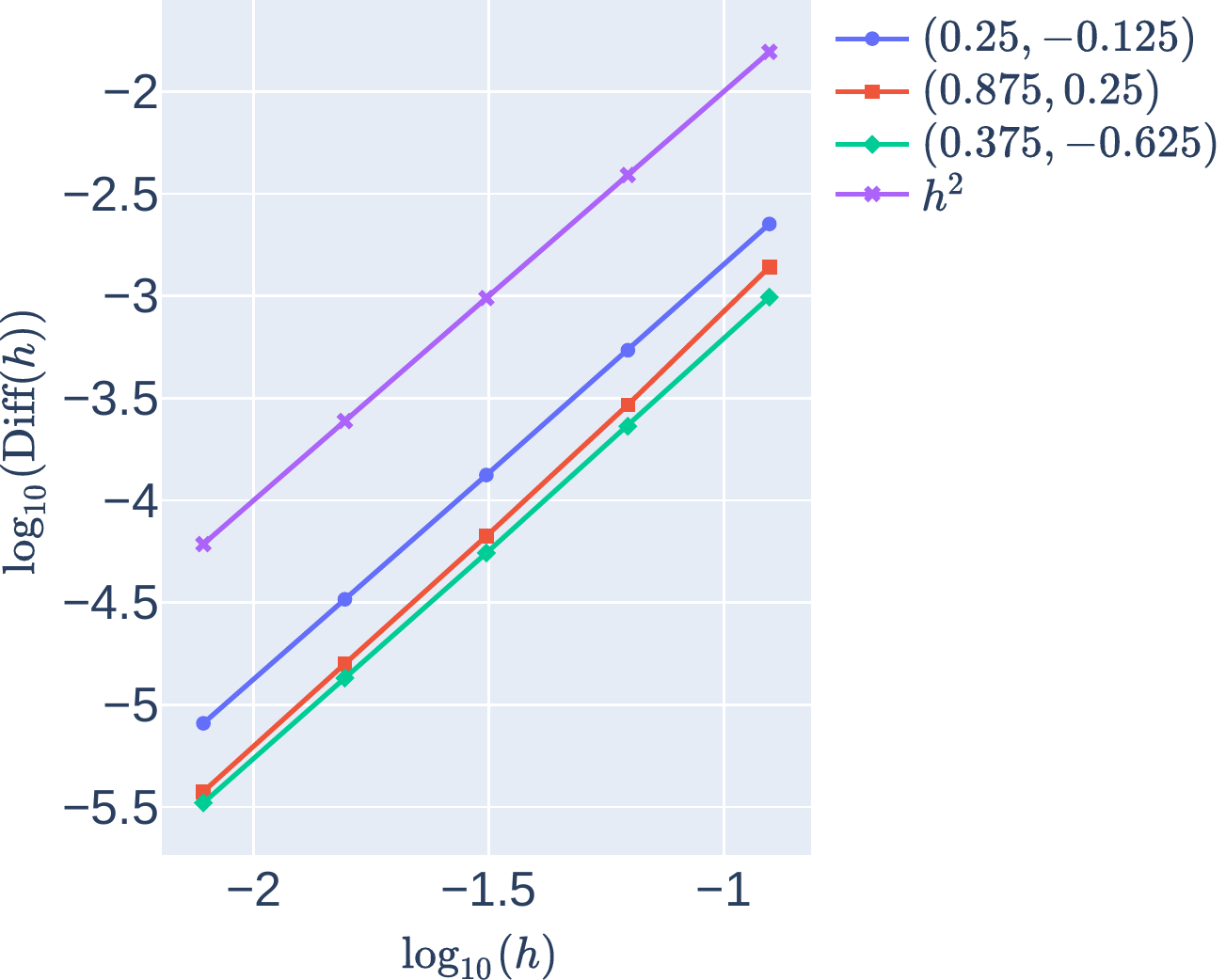}
         \caption{$\alpha = 1$ }
         \label{fig:alpha_one}
     \end{subfigure}
          \hfill
         \begin{subfigure}[b]{0.32\textwidth}
         \centering
         \includegraphics[width=\textwidth]{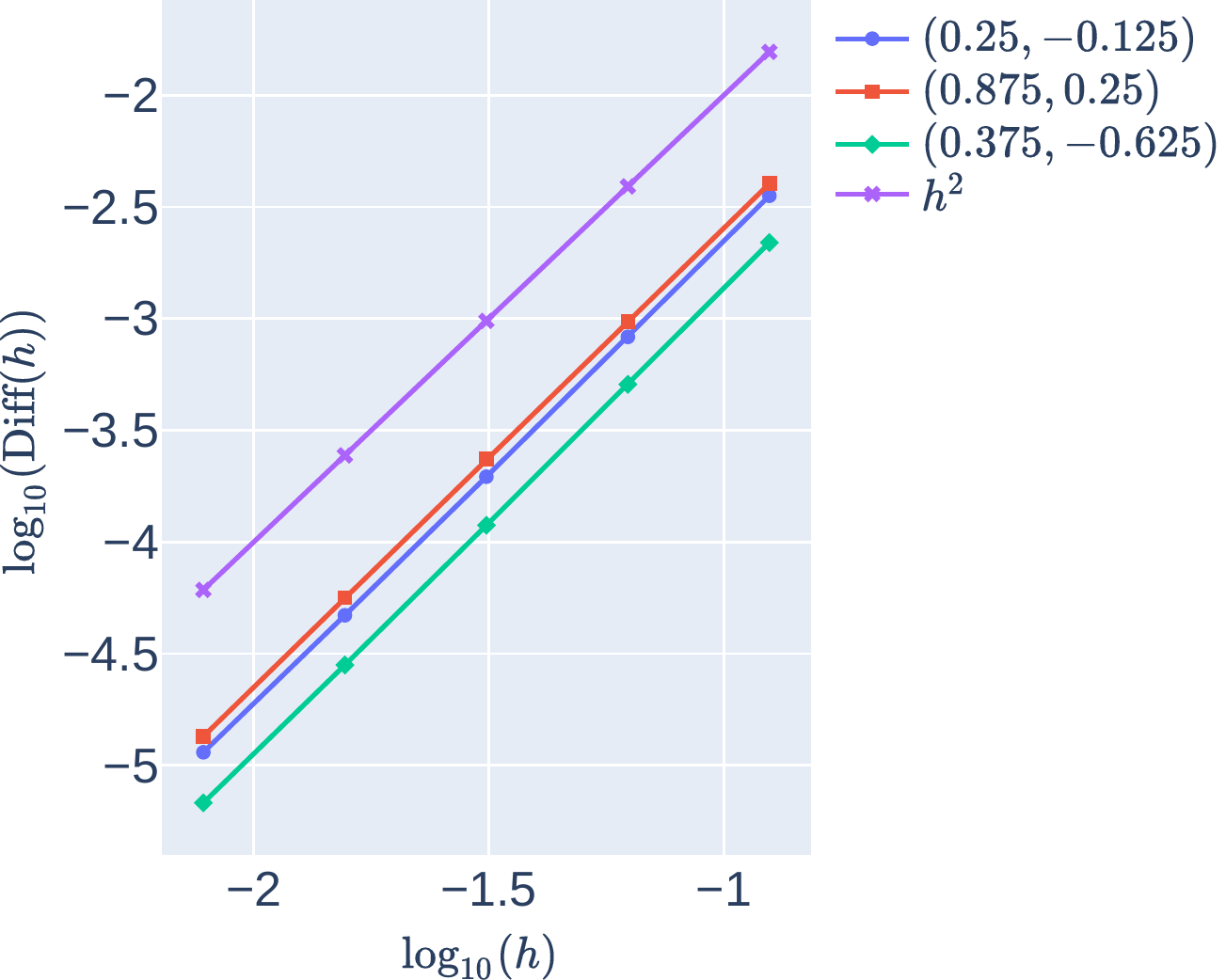}
         \caption{$\alpha = 1.5$ }
         \label{fig:alpha_one_half}
     \end{subfigure}
    \caption{The accuracy of the approximation $(-\Delta)_h^{\sfrac{h}{2}}$ for the fractional laplacian by \cref{eq:numerical_frac_lap}}
    \label{fig:order}
\end{figure}

We now verify the second-order accuracy of \cref{eq:numerical_frac_lap} in 2D, while the verification in other dimensions are similar. Let \begin{align}
    \phi(x)&\coloneqq (1 + x_1 + 2x_1^2)(1 + x_2^2)(1-x_1^2)_+^5(1-x_2^2)_+^5\in C_c^4(\R^2)
\end{align} so that $supp(\phi) = [-1,1]^2$. In 2D the correction weight $\omega_0$ is given by \begin{align}
    \omega_0 = \left\{\begin{array}{cc}
       0.960844610589965,  &  \alpha = 0.5\\
       1.950132460000978,  &  \alpha = 1 \\
       5.038779739396576, & \alpha = 1.5
    \end{array}\right. .
\end{align} We compute $(-\Delta)_h^{\sfrac{\alpha}{2}}\phi(x)$ at $x = (0.25, -0.125), (0.875, 0.25), (0.375, -0.625)$ for each $h = \frac{1}{2^3},\frac{1}{2^4}\cdots, \frac{1}{2^{12}}$. Since the analytical value of $(-\Delta)^{\sfrac{\alpha}{2}}\phi(x)$ is not available, we examine the absolute value of the difference \begin{align*}
    \text{Diff}(h) = |(-\Delta)_h^{\sfrac{\alpha}{2}}\phi(x) - (-\Delta)_{\frac{h}{2}}^{\sfrac{\alpha}{2}}\phi(x)|,
\end{align*} and the log-log plot between $\text{Diff}(h)$ and the mesh size $h$ to determine the order of accuracy. \cref{fig:order} confirms the second-order accuracy of $(-\Delta)_h^{\sfrac{\alpha}{2}}$ for $\alpha = 0.5, 1, 1.5$ respectively.

%%%%%%%%%%%%%%%%%%%%%%%%%%%%%%%%%%%%%%%%%%%%%%%%%%%%%%%%%%%%%%%%%%%%%%%%%%%%%%%%%%%%%%%%%%%%%%%%%%%%%%%%%%%%%%%%%%%%%%%%%%%%%%%%%%%%%%%%%%%%%%%%%%%%%%%%%%%%%%%%%%%%%%%%%%%%%%%%%%%%%%%%%%%%%%%%%%%%%%%%%%%%%%%%%%%%%%%%%%%%%%%%%%%%%%%%%%%%%%%%%%%%%%%%%%%%%%%%%%%%%%%%%%%%%%%%%%%%%%%%%%%%%%%%%%%%%%%%%%%%%%%%%%%%%%%%%%%%

\section{Structure of Physics-informed Neural Networks}\label{sec:PiNNs}
In this section we elaborate the structure of the trapz-PiNNs. It consists of three main parts: training data, neural network solver, loss function and training algorithm. \begin{description}
    \item[Training data:] We use the grid points $\mathcal{T}\coloneqq\{(x_j,t_k):1\leq j\leq J^n,1\leq k\leq K\}$, where $J = \frac{2}{h}$ and $K=\frac{T}{\Delta t}$, as training data, $(x_j)_j$, $(t_k)_k$ are uniform meshes for $\Omega$ and $(0,T)$ with mesh sizes $h$ and $\Delta t$, respectively. Denote $(x_j)_j$ by $\Omega_h$. Moreover, we can use validation data to monitor the training process and avoid overfitting phenomenon \cite{zhang2021dive} and the validation data can be set as $\mathcal{V}\coloneqq \Omega_h\times \{T_{\text{pred}}\}$.
    \item[Neural Network solver:] We use the fully-connected feedforward neural network (NN) in this paper, which is the foundation for all variants of neural networks \cite{zhang2021dive}. It is also the core part of our NN solver \cref{eq:approx sol}. A NN $u_{nn}$ with input dimension $d\in\N$, depth $N+1\in\N$ with $N$'s hidden layers, width $M\in\N$ and output dimension $d'\in\N$ is of the form \begin{align}
    u_{nn}(x;\Theta) = A_{N+1} \circ \sigma \circ A_{N} \circ \sigma \circ A_{N-1} \circ \cdots \circ \sigma \circ A_1(x), \label{eq:NN}
            \end{align} where \begin{enumerate}
        \item $A_1:\R^d\rightarrow \R^M$, $A_i:\R^M\rightarrow \R^M,\ 2\leq i\leq N$ and $A_{N+1}:\R^M \rightarrow \R^{d'}$ are affine transformations with matrix representations \begin{align}
        A_i(x) &= W_ix + b_i,\quad i = 1,\dots,N+1, \nonumber 
        \end{align} 
        \item $\sigma:\R\rightarrow\R$ is a non-linear activation function. With some abuse of notation, we follow the convention that $\sigma$ has the vectorized form $\sigma(x) = (\sigma(x_1),\cdots, \sigma(x_n)),\ \forall x\in \R^n,\ \forall n\in\N$. 
        \item The (learnable) parameters $\Theta$ of NN in \cref{eq:NN} are \begin{align}
        \Theta \coloneqq \{(W_i,b_i):i = 1,\cdots, N+1\} \cong \R^{(N-1)M^2+(N+d+d')M+d'}.
        \end{align}
        \end{enumerate} We call an NN has \emph{shape} $(M, N+1)$ if NN has width $M$ and depth $N+1$. To solve the FPE \crefrange{eq:Fokker-Planck}{eq:init cond}, it is convenient to write the NN solver as an ansatz form \begin{align}
                 \hat{u}(x,t;\Theta) = \rho(t) \tau(x) u_{nn}(x,t;\Theta) + u_0(x), \label{eq:approx sol}
            \end{align} so that $\hat{u}$ satisfies the initial-boundary condition \cref{eq:init cond,,eq:bdry cond}. Here, $\rho(t)$ is a non-negative, bounded, strictly-increasing $C^\infty([0,\infty))$ function such that $\rho(0) = 0$ and $\tau(x)$ is a $C^\infty(\Omega)$ function with $\tau|_{\Omega^c} = 0$. The input and output dimensions of neural network $u_{nn}(t,x;\Theta)$ are $n+1$ and $1$ while the width and depth vary. A good choice of functions $\rho(t),\ \tau(x)$ deserves careful consideration. We give a heuristic explanation here. The Universal Approximation Theorem \cite{HORNIK1991251,kidger2020universal} guarantees that NN with arbitrary width bounded depth or bounded width arbitrary depth uniformly converges to continuous function on compact set in $\R^n$ up to arbitrary degree of precision. By the ansatz form \cref{eq:approx sol}, the NN tries to learn \begin{align}
                u_{nn}(t,x;\Theta) \underset{\Theta}{\longrightarrow} \frac{u(t,x) - u_0(x)}{\rho(t)\tau(x)}\eqqcolon \varphi(t,x),
            \end{align} where we assume $u\in C(\Omega_T)$ is the true solution and $u_0$ is the initial condition. Since $\Omega_T$ is open, we prefer $\varphi$ can be continuously extended to the compact set $\overline{\Omega_T}$ so that $\varphi\in C(\overline{\Omega_T})$. This is possible if and only if $\rho$ and $\tau$ are chosen such that $\varphi$ has well-defined limiting behaviors as $x\to\partial\Omega$ and/or $t\to 0+$.
    \item[Loss function and training:] The loss function $\mathbf{L}(\Theta)$ is the classic Mean Squared Error (MSE). We adopt the mini-batch training in this work, the loss function is therefore evaluated on a randomly selected training batch $\mathcal{B}\subset\mathcal{T}$ of fixed size in each epoch, i.e. \begin{align}
            \mathbf{L}_{\mathcal{B}}(\Theta) = \frac{1}{|\mathcal{B}|}\sum_{(x_j,t_k)\in\mathcal{B}}|\partial_t\hat{u}(x_j,t_k;\Theta) - \mathcal{L}_h\hat{u}(x_j,t_k;\Theta)|^2. \label{eq:loss_function}
        \end{align} $\mathcal{L}_h$ denotes the discretized operator of $\mathcal{L}$ at \cref{eq:Fokker-Planck}. In the RHS of \cref{eq:loss_function}, (partial) derivatives with integer order can be computed directly by automatic differentiation \cite{griewank2008evaluating}, fractional laplacian is evaluated by modified trapezoidal rule \cref{eq:numerical_frac_lap}. The grid points for the trapezoidal rule coincide with training data $\mathcal{T}$. The NN solver at \cref{eq:approx sol} already satisfied the initial and boundary conditions. Otherwise, proper loss functions with respect to initial and boundary conditions have to be added into \cref{eq:loss_function}. Once we set up the training data, NN solver and loss function, we are looking for best possible parameters $\Theta^*$ that minimize the loss function, i.e. \begin{align}
            \Theta^* = \underset{\Theta}{\arg\min}\ \mathbf{L}(\Theta).
        \end{align} To achieve this, we use ADAM, a popular stochastic gradient descent (SGD) algorithm \cite{kingma2014adam}. Several hyperparameters are required to be initialized before running the algorithm. They are NN shape, batch size, learning rate, epoch, time and space resolutions. In addition, a initial guess $\Theta_0$ is provided by Xavier Initialization scheme \cite{glorot2010understanding}.
\end{description} We summarize this section into the following \cref{alg:cap}.
\begin{algorithm}
\caption{Training algorithm for trapz-PiNN}\label{alg:cap}
\begin{algorithmic}
\Require NN shape, batch size $B$, learning rate $\eta$, epoch $K$, space resolution $h$, time resolution $\Delta t$ and initial parameter $\Theta_0$.
\Ensure 
\State Initialize $\hat{u}(\cdot;\Theta_0)$, $\mathcal{T}$ and $\mathcal{V}$.
\For {$i = 0,\cdots, K-1$}
\State 1. Randomly select $\mathcal{B}\in\mathcal{T}$ with $|\mathcal{B}| = B$
\State 2. Compute \begin{align}
    \nabla_{\Theta}\mathbf{L}_{\mathcal{B}}(\Theta)\vert_{\Theta = \Theta_i} = \nabla_{\Theta}\bigg\{ \frac{1}{|\mathcal{B}|}\sum_{(x_j,t_k)\in\mathcal{B}}|\partial_t\hat{u}(x_j,t_k;\Theta) - \mathcal{L}_h\hat{u}(x_j,t_k;\Theta)|^2 \bigg\}\bigg\vert_{\Theta = \Theta_i} \nonumber
\end{align}
\State 3. Update $\Theta_{i+1} = \Theta_i -  \eta \nabla_{\Theta}\mathbf{L}_{\mathcal{B}}(\Theta)\vert_{\Theta = \Theta_{i}}$ \Comment{This is vanilla SGD, we use ADAM in practice}
\State 4. Randomly select $\mathcal{B}'\in\mathcal{V}$ with $|\mathcal{B}'| = B$
\State 5. Record the training error $\mathbf{L}_{\mathcal{B}}(\Theta_{i+1})$ and the validation error $\mathbf{L}_{\mathcal{B}'}(\Theta_{i+1})$. \Comment{For monitoring the training process}

\EndFor
\end{algorithmic}
\end{algorithm}

%%%%%%%%%%%%%%%%%%%%%%%%%%%%%%%%%%%%%%%%%%%%%%%%%%%%%%%%%%%%%%%%%%%%%%%%%%%%%%%%%%%%%%%%%%%%%%%%%%%%%%%%%%%%%%%%%%%%%%%%%%%%%%%%%%%%%%%%%%%%%%%%%%%%%%%%%%%%%%%%%%%%%%%%%%%%%%%%%%%%%%%%%%%%%%%%%%%%%%%%%%%%%%%%%%%%%%%%%%%%%%%%%%%%%%%%%%%%%%%%%%%%%%%%%%%%%%%%%%%%%%%%%%%%%%%%%%%%%%%%%%%%%%%%%%%%%%%%%%%%%%%%%%%%%%%%%%%%%%%%%%%%%%%%%%%%%%%%%%%%%%%%%%%%%%%%%%%%%%%%%%%%%%%%%%%%%%%%%%%%%%%%%%%%%%%%%%%%%%%%%%%%%%%%%%%%%%%%%%%%%%%%%%%%%%%%%%%%%%%%%%%%%%%%%%%%%%%%%%%%%%%%%%%%%%%%%%%%%%%%%%%%%%%%%%%%%%%%%%%%%%%%%%%%%%%%%%%%%%%%%%%%%%%%%%%%%%%%%%%

\section{Numerical Experiments}\label{sec:num experiment}
In this section we demonstrate the expressive power of trapz-PiNNs. In each numerical examples, the trapz-PiNNs are trained on $\Omega_{T}$ with $T = 0.2$ for each equation and predict the solutions at multiple time steps starting from $t = 0.2$. Moreover, we verify the accuracy of DL solution by comparing with the reference solutions obtained from FDM together with modified trapezoidal rule \cref{eq:modified_trapz} and determining the $\mathcal{L}^2$ relative error. We also investigate the DL solution profile from the pointwise absolute and relative errors. 

Let $\hat{u},u_{\text{ref}}$ be the DL solution and reference solution, respectively. Recall that $\Omega_h = (x_j)_j$ is uniform mesh for $\Omega$ with mesh size $h$. Given a sample time $t$, the $\mathcal{L}^2$ relative error at $t$ is defined by \begin{align} \label{eq:veps}
    \varepsilon = \sqrt{\frac{\sum_{j}|\hat{u}(x_j,t)-u_{\text{ref}}(x_j,t)|^2}{\sum_{j}|u_{\text{ref}}(x_j,t)|^2}}
\end{align} and for each $x_j\in \Omega_h$, the absolute and relative errors at $x_j$ are defined by \begin{align}
    \epsilon_{\text{abs}}(x_j) &= |\hat{u}(x_j,t)-u_{\text{ref}}(x_j,t)|, \label{eq:eps_abs}\\
    \epsilon_{\text{re}}(x_j) &= \left\vert\frac{\epsilon_{\text{abs}}(x_j,t)}{u_{\text{ref}}(x_j,t)}\right\vert. \label{eq:eps_re}
\end{align} The activation function $\sigma$, the functions $\rho(t)$ and $\tau(x)$ in the ansatz (\ref{eq:approx sol}) are fixed throughout  \begin{align}
    \sigma(x) &= \frac{1}{1 + \exp(-x)},\quad \text{sigmoid function} \\
    \rho(t) &= 1 - \exp(-t),\\
    \tau(x) &= \left(\prod_{k=1}^n(1-x_k^2)_+\right)^{\frac{1}{2}},
\end{align} and the hyperparameters will be specified in each example. It worths mentioning that $\tau(x)$ is chosen a posterior to the initial condition.

To construct FDM reference solution for FPE \crefrange{eq:Fokker-Planck}{eq:init cond}, we use grid points $\mathcal{T}$ where, if not specified, space resolution $h = \frac{1}{32}$ and time step $\Delta t = \frac{h^2}{4}$. If any, first-order partial derivatives $\pa_{x_j}\{\cdot\}, j = 1,\cdots,n$ are evaluated by classical upwind scheme, second-order partial derivatives $\pa_{x_ix_j}^2\{\cdot\}, i,j = 1,\cdots,n$ are discretized by central differences. Fractional laplacian is discretized by modified trapezoidal rule \cref{eq:modified_trapz}. For time evolution, we adopt third-order total-variation diminishing Runge-Kutta scheme, i.e., given an ODE $\frac{du}{dt} = R(u)$, \begin{align*}
    U^{(1)} &= U^{n} + \Delta t R(U^{n}), \\
    U^{(2)} &= \frac{3}{4}U^{n} + \frac{1}{4}U^{(1)} + \frac{1}{4}\Delta t R(U^{(1)}), \\
    U^{(3)} &= \frac{1}{3}U^{n} + \frac{2}{3}U^{(2)} + \frac{2}{3}\Delta t R(U^{(2)}),
\end{align*} where $U^n$ denotes numerical solution of $u$ evaluated at time $t = t_n$.

As a common practice, we adopt $\mathcal{L}^2$ relative error $\varepsilon = 0.01$ as benchmark. DL solution with $\mathcal{L}^2$ relative error $\veps<0.01$ is considered accurate in general. We call this accuracy by $\varepsilon$-accuracy. Another standard is attaining relative error $\epsilon_{\text{re}}<1\%$ uniformly over domain. We stress that this is a very strict critera for DL solvers in general and is hard to achieve in reality. One primary reason is that the DL solvers are optimized with respect to MSE, which itself is an indicator of global accuracy. To author's best knowledge, without utilizing external information such as incorporating high-fidelty simulated solution into loss function, there is no known loss function aiming for such high accuracy of (maximum) relative error.

All numerical experiments are coded in Python with machine learning package JAX \cite{jax2018github} and performed on a Google Colab Pro+ account and the source codes are available at \url{https://github.com/sjiang23}.
%%%%%%%%%%%%%%%%%%%%%%%%%%%%%%%%%%%%%%%%%%%%%%%%%%%%%%%%%%%%%%%%%%%%%%%%%%%%%%%%%%%%%%%%%%%%%%%%%%%%%%%%%%%%%%%%%%%%%%%%%%%%%%%%%%%%%%%%%%%%%%%%%%%%%%%%%%%%%%
\subsection{Fractional Heat Equation}\label{subsection:frac_heat_2D}
In both 2D and 3D, we consider the following fractional heat equation with the initial and boundary conditions: \begin{align}
    \partial_tu(x,t) &= -(-\Delta)^{\frac{\alpha}{2}}u(x,t), \label{eq:heat}\\
    u(x,0) &= c_n \left(\prod_{j=1}^n(1-x_j^2)_+\right)^4, \label{eq:heat init cond}\\
    u(x,t) &= 0,\quad x\not \in (-1,1)^n.\label{eq:heat bdry}
\end{align} Here $n = 2,3$, $c_2 = (\frac{315}{216})^2$ and $c_3 = (\frac{693}{512})^3$, where $c_n$ are chosen so that $\int_{\R^n}u(x,0) \d x = 1$. It is not necessary to choose $u(x,0)$ to be probability density and other choices are also feasible. We solve \crefrange{eq:heat}{eq:heat bdry} for $\alpha = 0.5, 1, 1.5$ in 2D and $\alpha = 1$ in 3D.
%%%%%%%%%%%%%%%%%%%%%%%%%%%%%%%%%%%%%%%%%%%%%%%%%%%%%%%%%%%%%%%%%%%%%%%%%%%%%%%%%%%%%%%%%%%%%%%%%%%%%%%%%%%%%%%%%%%%%%%%%%%%%%%%%%%%%%%%%%%%%%%%%%%%%%%%%%%%%% subsubsection heat 2D
\subsubsection{2D case} \label{subsubsection:heat 2D}
\begin{figure}[ht]
    \begin{subfigure}[t]{0.5\textwidth}
        \includegraphics[width = \textwidth]{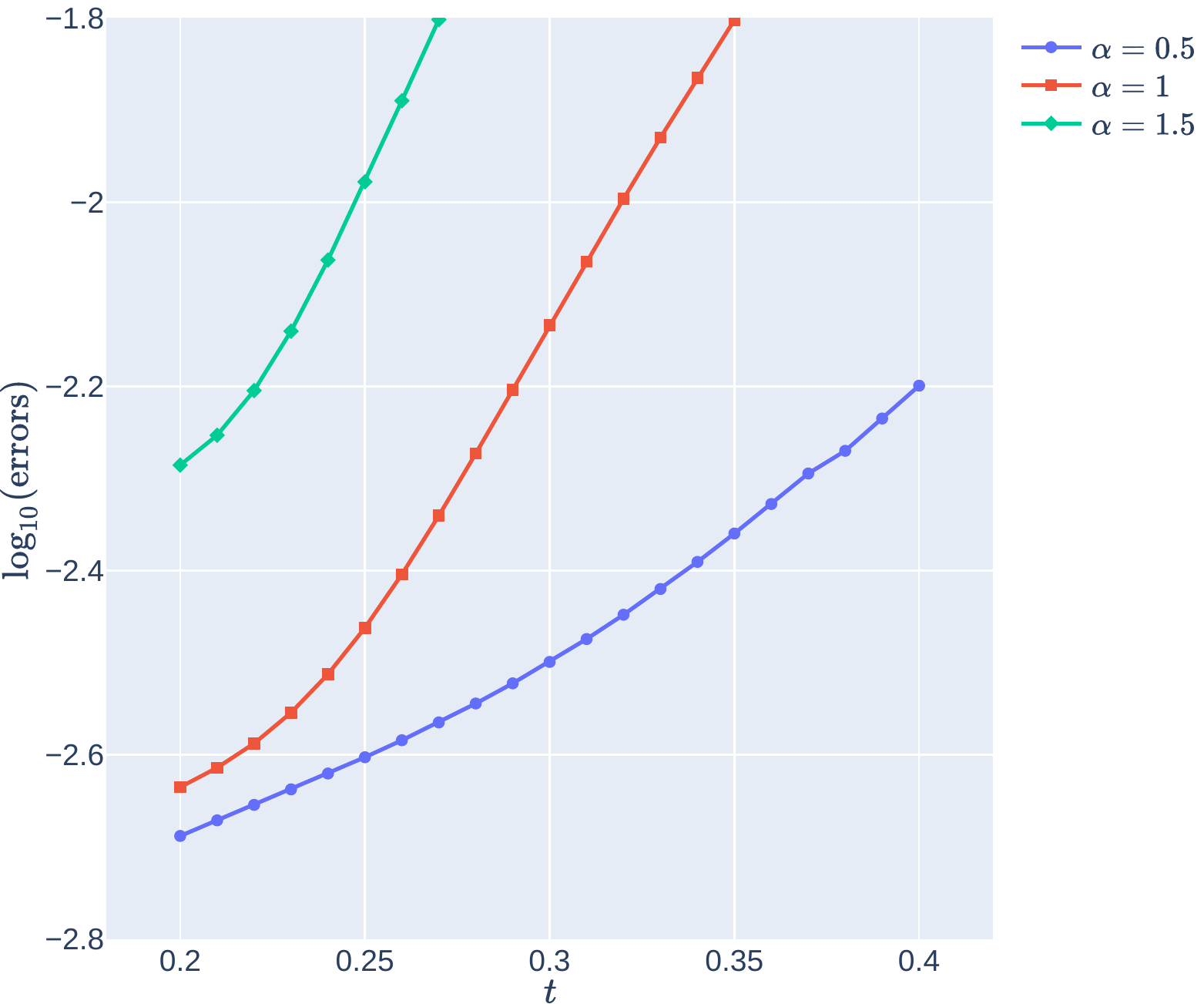}
         \caption{Fractional heat equation in 2D}
        \label{fig:heat_2D_range_of_good_accuracy}
    \end{subfigure}
    \hfill
    \begin{subfigure}[t]{0.5\textwidth}
        \includegraphics[width = \textwidth]{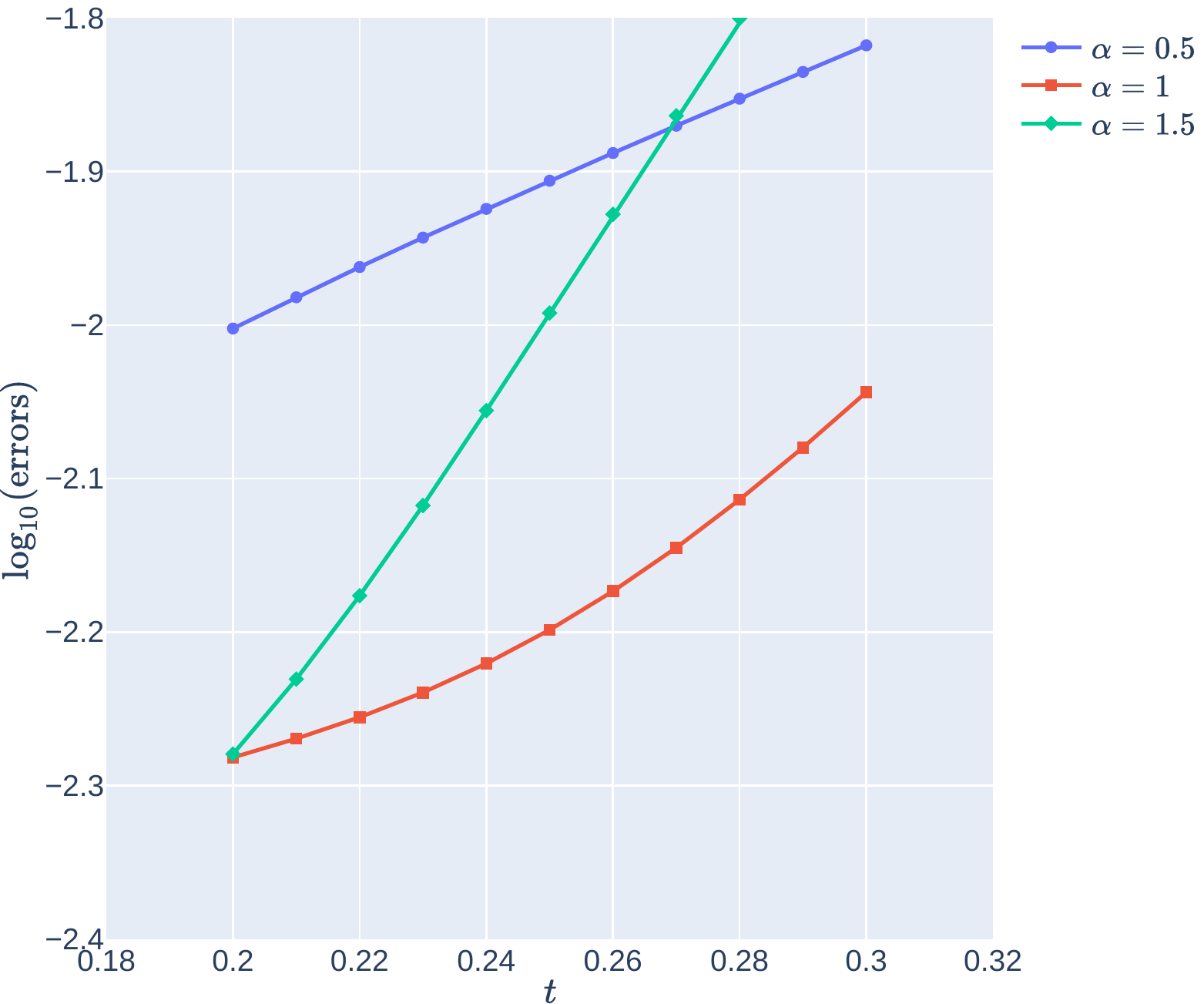}
         \caption{Fokker-Planck equation with O-U potential in 2D}
        \label{fig:OU_range_of_good_accuracy}
    \end{subfigure}
    \caption{The $\mathcal{L}^2$ relative errors $\veps$ at time steps $0.2 + \sfrac{k}{100}, k = 0,\cdots,25$ for the DL solutions of (a) fractional heat equations in 2D (\crefrange{eq:heat}{eq:heat bdry}) and (b) FPEs with O-U potential in 2D (\crefrange{eq:OU}{eq:OU bdry}) when $\alpha = 0.5, 1, 1.5$.}
    \label{fig:range_of_good_accuracy}
\end{figure}

\begin{figure}[ht]
    \begin{subfigure}[t]{0.24\textwidth}
         \includegraphics[width=\textwidth]{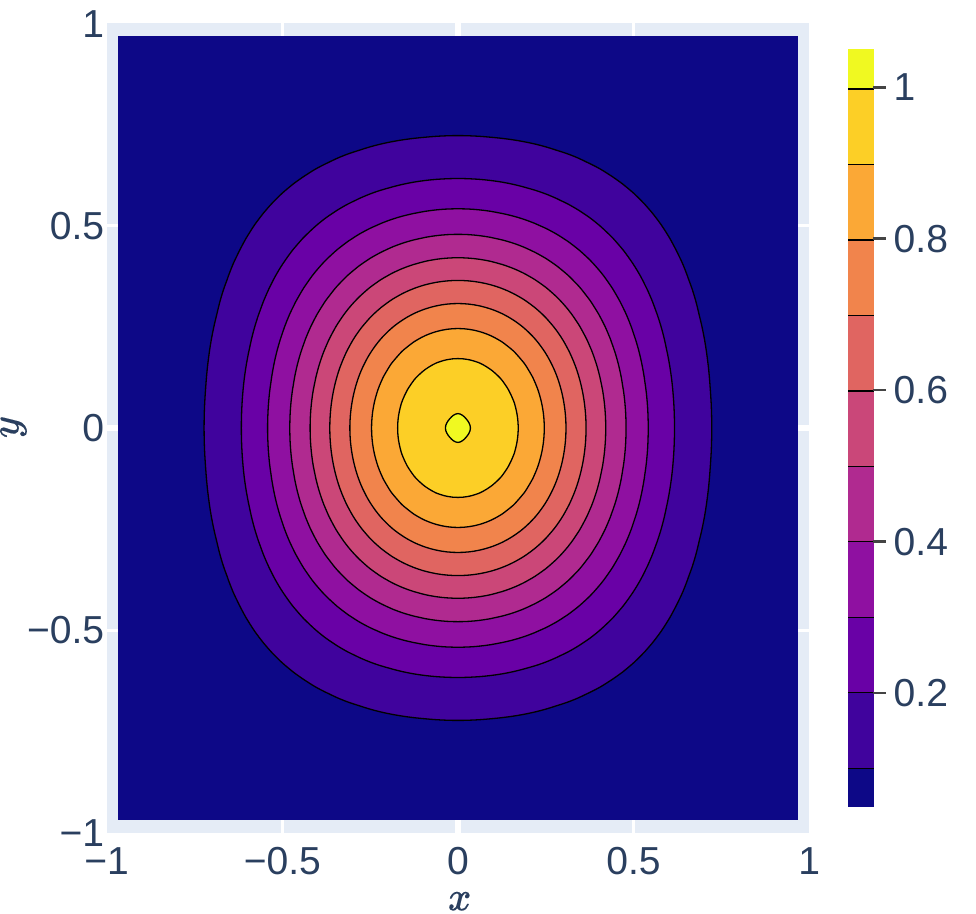}
         \caption{FDM solution \\ $\alpha = 0.5$}
         \label{fig:FDM_alpha_0.5}
     \end{subfigure}
     \hfill
         \begin{subfigure}[t]{0.24\textwidth}
         \includegraphics[width=\textwidth]{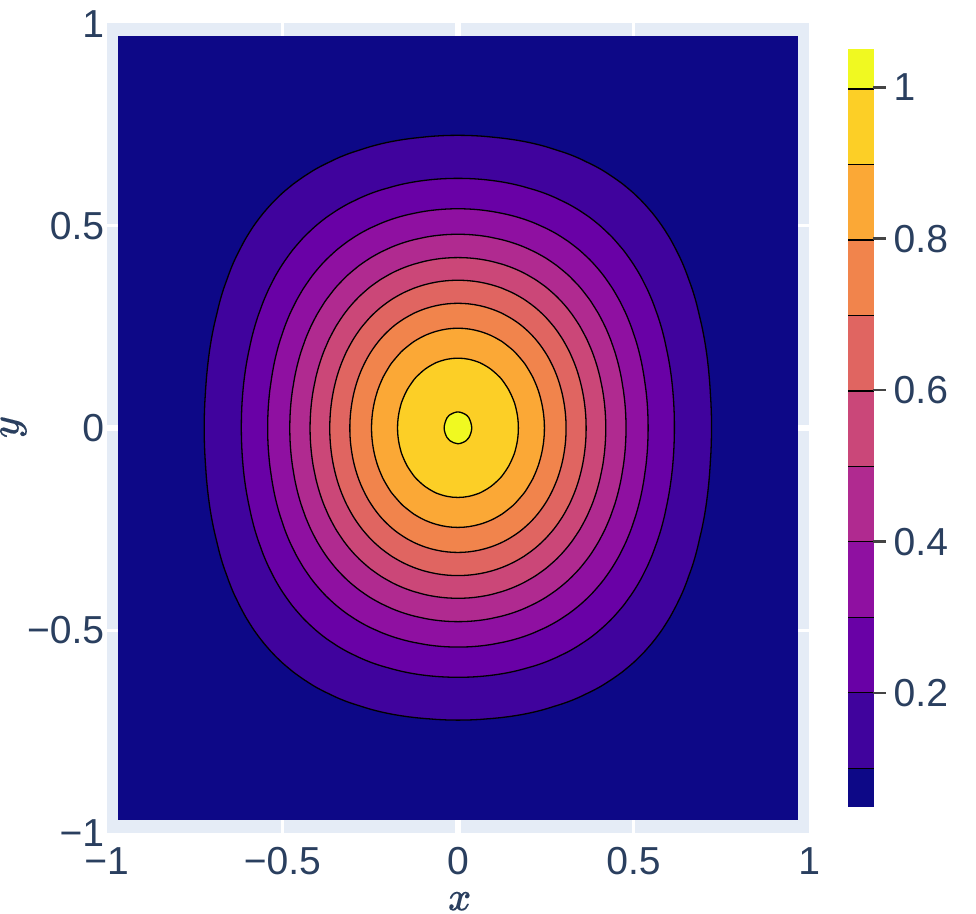}
         \caption{DL solution \\ $\alpha = 0.5$}
         \label{fig:NN_alpha_0.5}
     \end{subfigure}
      \hfill
         \begin{subfigure}[t]{0.24\textwidth}
         \includegraphics[width=\textwidth]{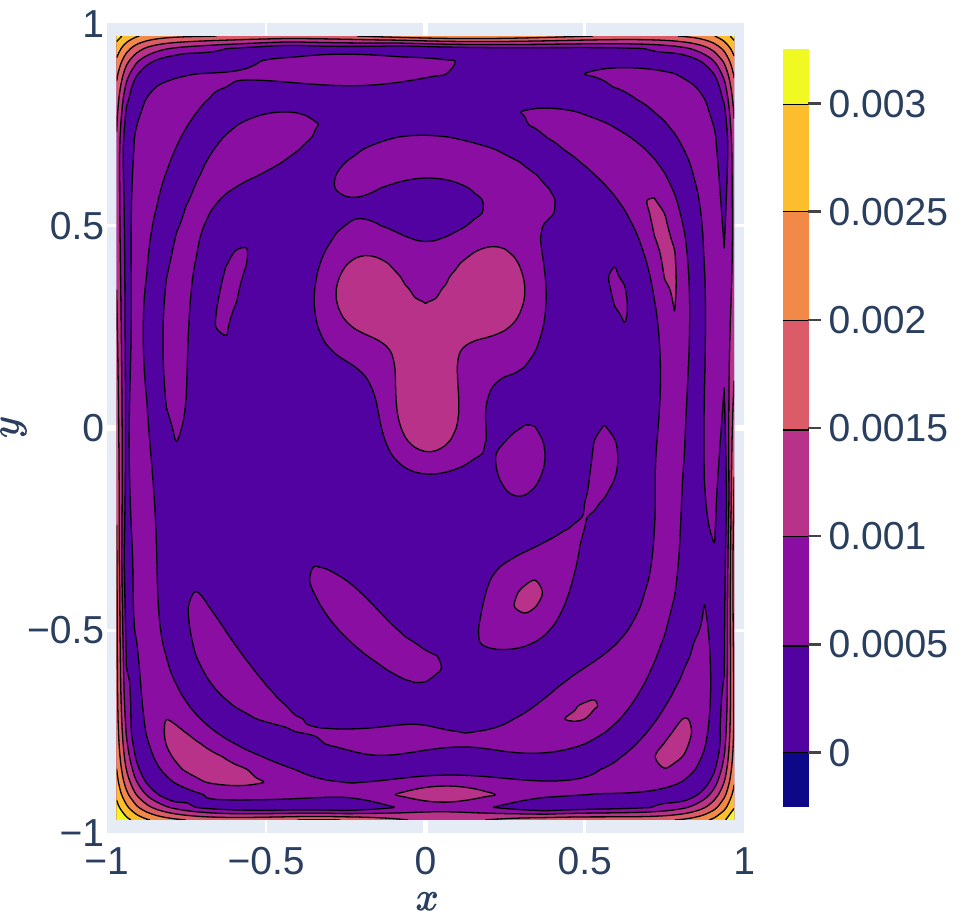}
         \caption{$\eps_{\text{abs}}$ for $\alpha = 0.5$}
         \label{fig:abs_err_alpha_0.5}
     \end{subfigure}
        \hfill
         \begin{subfigure}[t]{0.24\textwidth}
         \includegraphics[width=\textwidth]{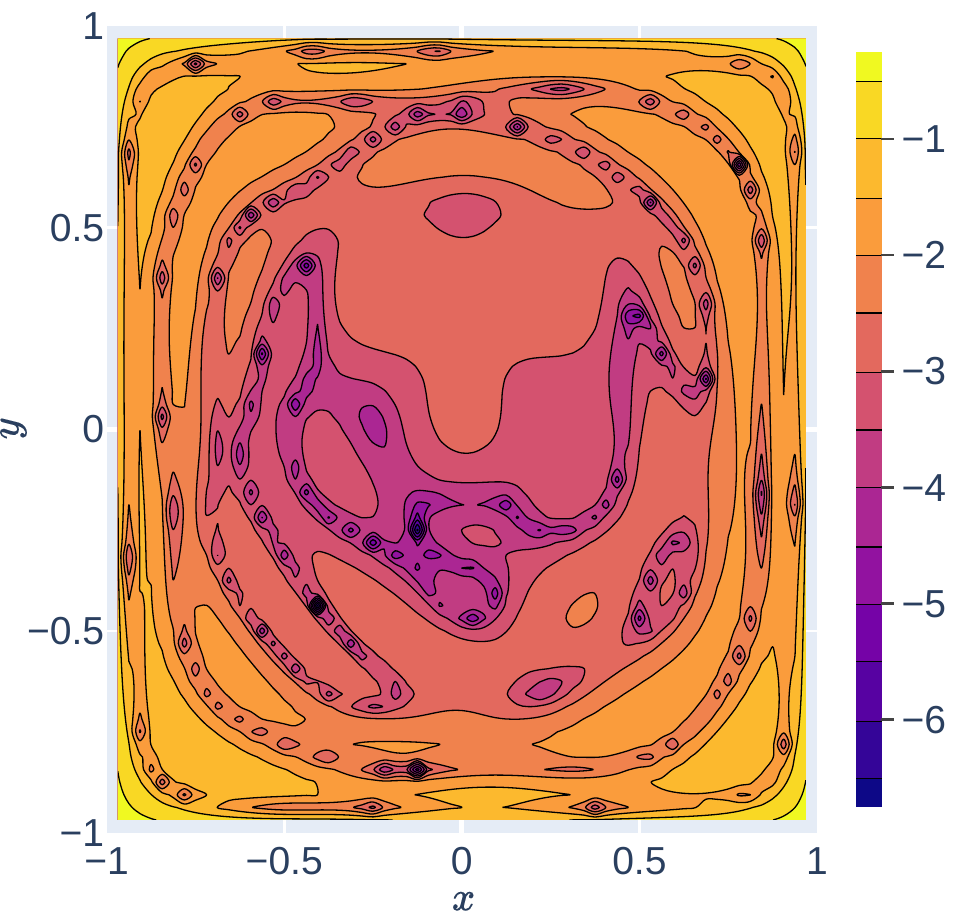}
         \caption{$\log_{10}(\epsilon_{\text{re}})$ for $\alpha = 0.5$}
         \label{fig:r_err_alpha_0.5}
     \end{subfigure}
     \hfill
         \begin{subfigure}[t]{0.24\textwidth}
         \includegraphics[width=\textwidth]{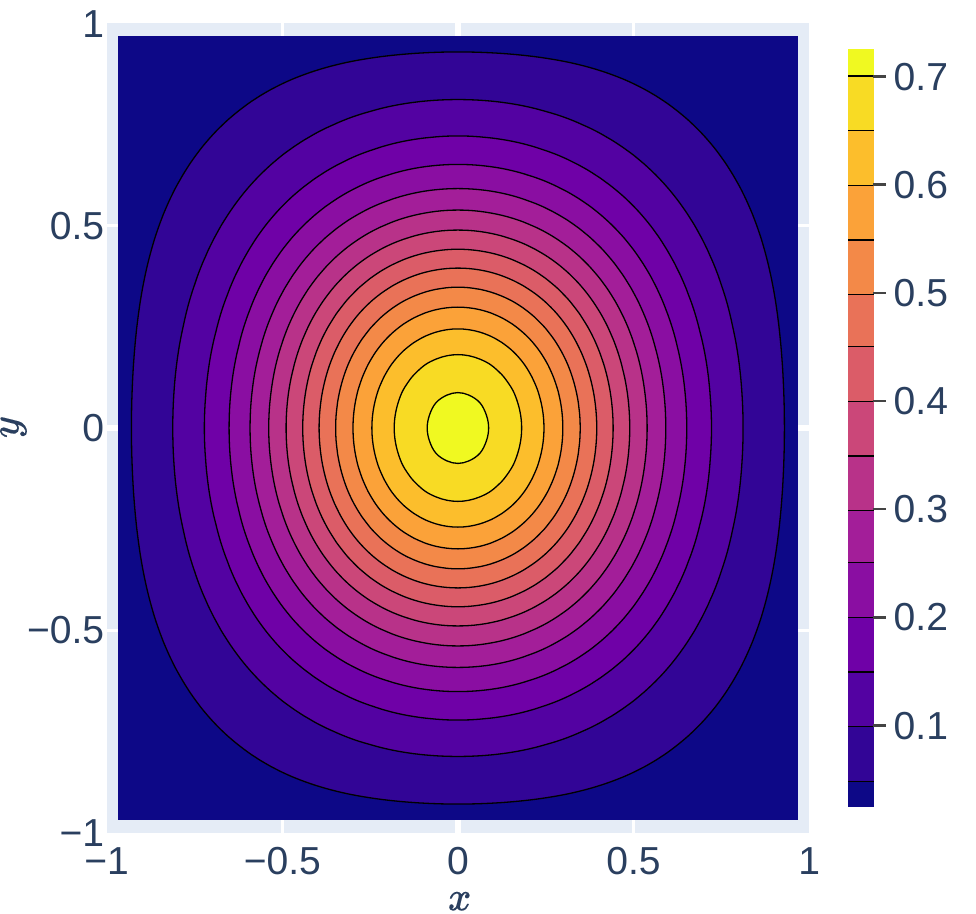}
         \caption{FDM solution \\ $\alpha = 1$}
         \label{fig:FDM_alpha_1}
     \end{subfigure}
     \hfill
         \begin{subfigure}[t]{0.24\textwidth}
         \includegraphics[width=\textwidth]{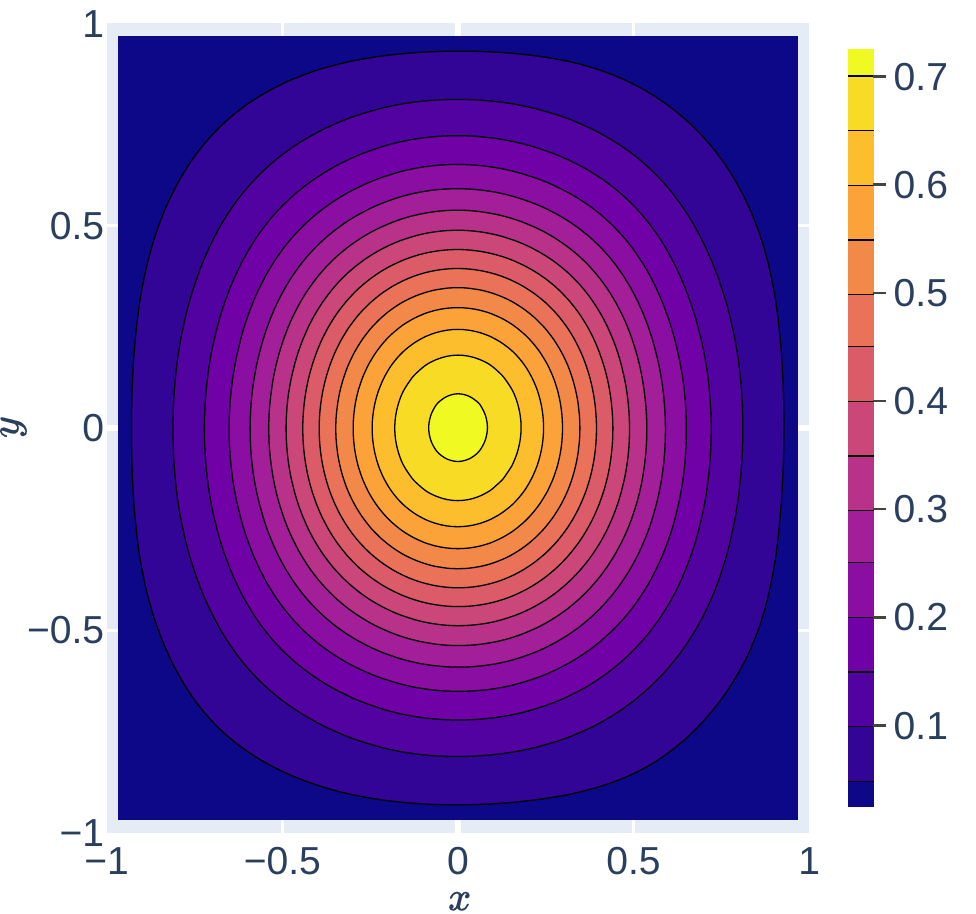}
         \caption{DL solution \\ $\alpha = 1$}
         \label{fig:NN_alpha_1}
     \end{subfigure}
      \hfill
         \begin{subfigure}[t]{0.24\textwidth}
         \includegraphics[width=\textwidth]{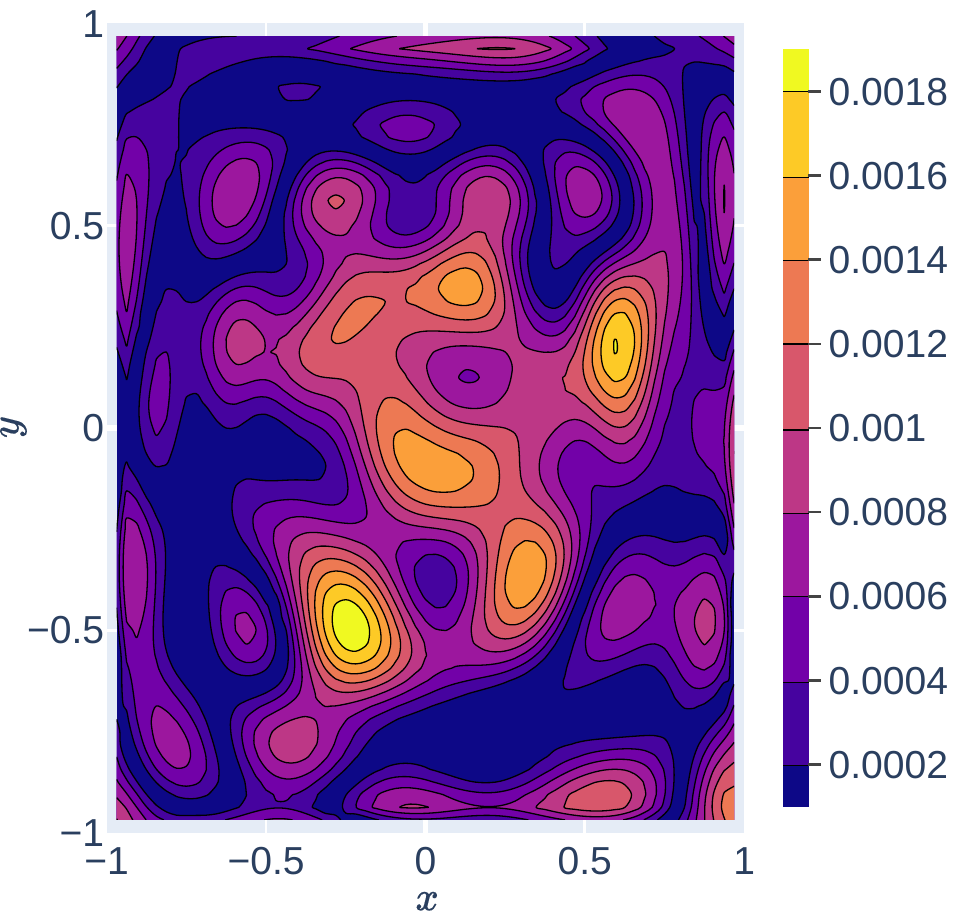}
         \caption{$\eps_{\text{abs}}$ for $\alpha = 1$}
         \label{fig:abs_err_alpha_1}
     \end{subfigure}
        \hfill
         \begin{subfigure}[t]{0.24\textwidth}
         \includegraphics[width=\textwidth]{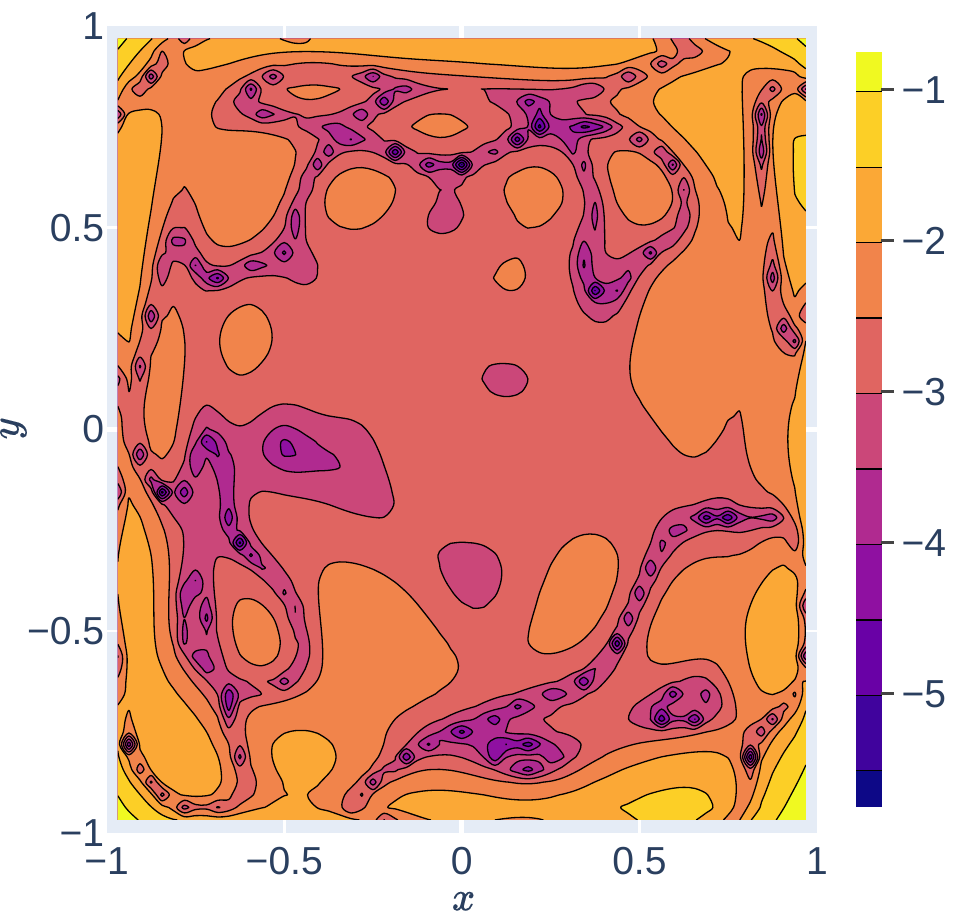}
         \caption{$\log_{10}(\epsilon_{\text{re}})$ for $\alpha = 1$}
         \label{fig:r_err_alpha_1}
     \end{subfigure}
     \hfill
         \begin{subfigure}[t]{0.24\textwidth}
         \includegraphics[width=\textwidth]{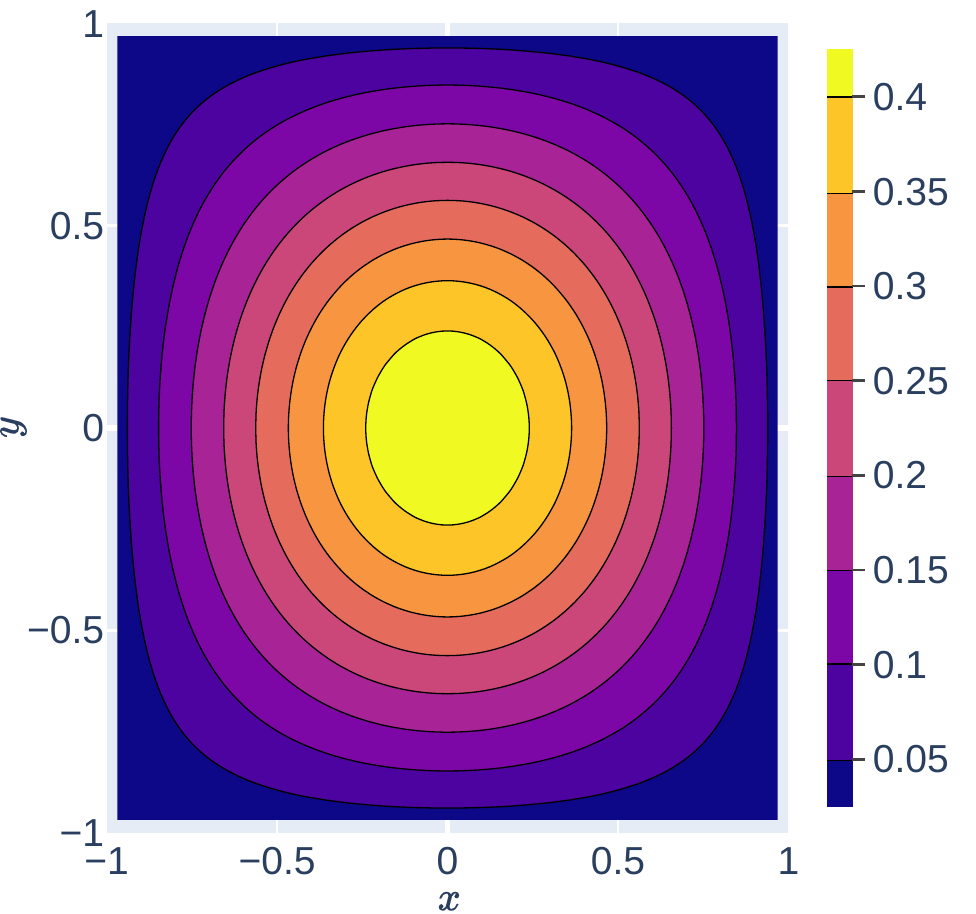}
         \caption{FDM solution \\ $\alpha = 1.5$}
         \label{fig:FDM_alpha_1.5}
     \end{subfigure}
     \hfill
         \begin{subfigure}[t]{0.24\textwidth}
         \includegraphics[width=\textwidth]{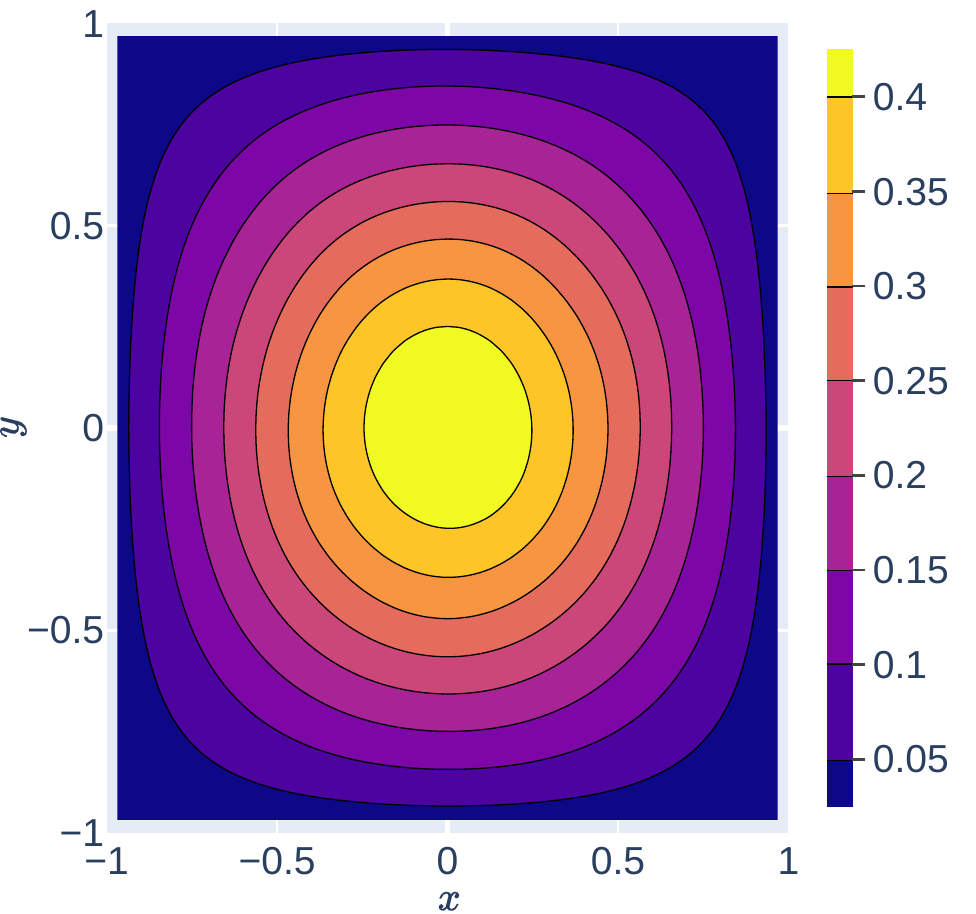}
         \caption{DL solution \\ $\alpha = 1.5$}
         \label{fig:NN_alpha_1.5}
     \end{subfigure}
        \hfill
         \begin{subfigure}[t]{0.24\textwidth}\centering
         \includegraphics[width=\textwidth]{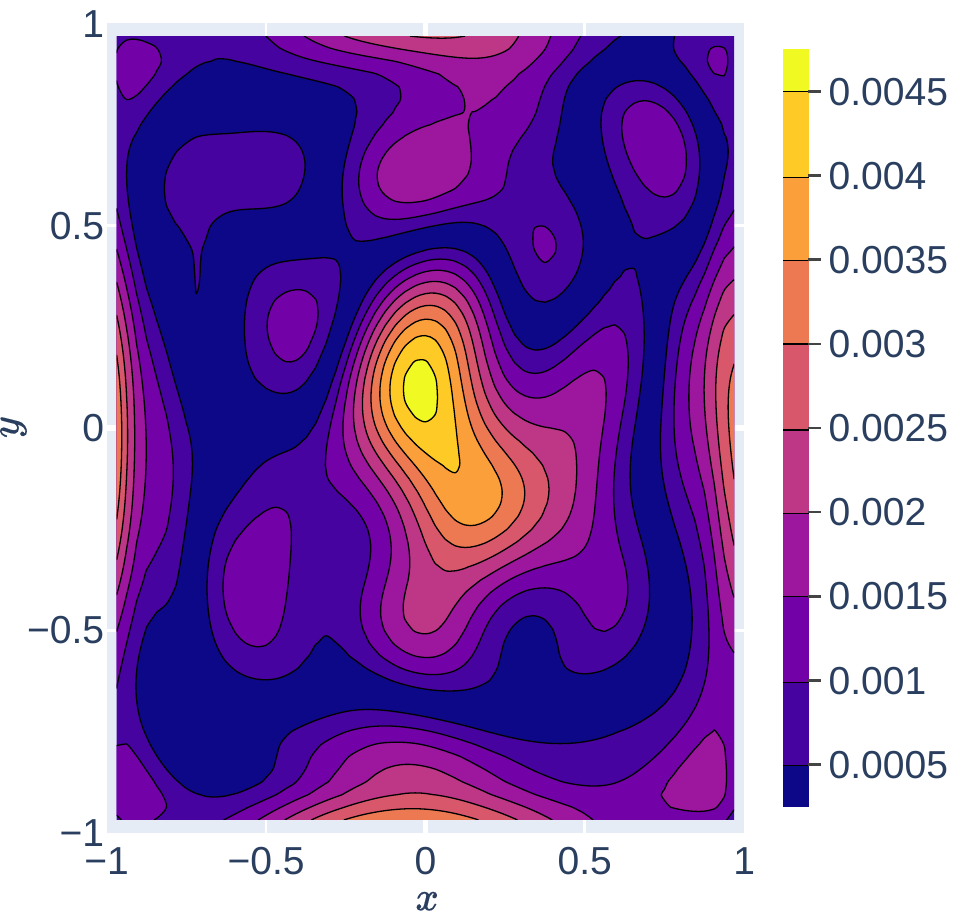}
         \caption{$\eps_{\text{abs}}$ for $\alpha = 1.5$}
         \label{fig:abs_err_alpha_1.5}
     \end{subfigure}
        \hfill
         \begin{subfigure}[t]{0.24\textwidth}
         \includegraphics[width=\textwidth]{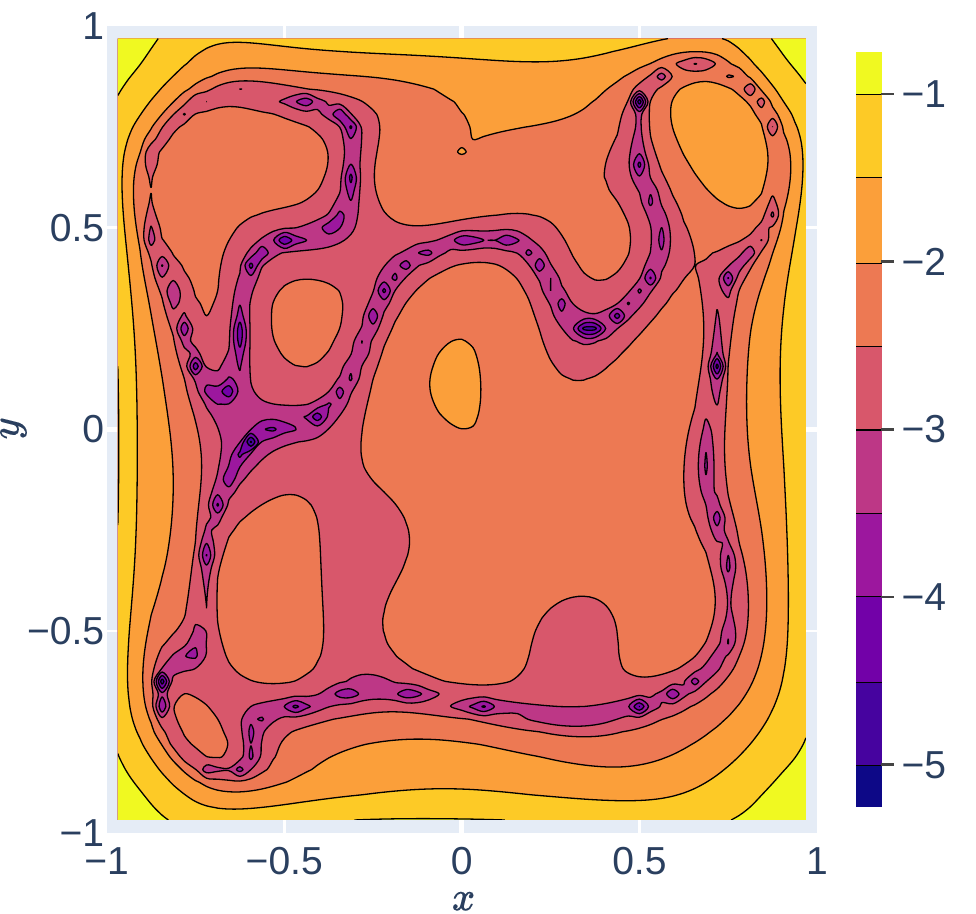}
         \caption{$\log_{10}(\epsilon_{\text{re}})$ for $\alpha = 1.5$}
         \label{fig:r_err_alpha_1.5}
     \end{subfigure}
    \caption{The contour plots of the FDM solutions, the DL solutions, the absolute and relative errors of the DL solutions for the fractional heat equations in 2D \crefrange{eq:heat}{eq:heat bdry} sampled at $t = 0.225$ for $\alpha = 0.5, 1, 1.5$. }
    \label{fig:heat_2D}
\end{figure}

In 2D, the hyperparameters for trapz-PiNNs are chosen as the NN shape $(20, 5)$, the learning rate $10^{-3}$, the batch size $64$, the epoch $2\times 10^5$, the space resolution $\sfrac{1}{32}$ and the time resolution $10^{-2}$. \cref{fig:heat_2D_range_of_good_accuracy} records the $\mathcal{L}^2$ relative errors $\veps$ defined in \cref{eq:veps} at times steps $T + \sfrac{k}{100}, k = 0,\cdots,25$. \cref{fig:heat_2D} presents the FDM reference solutions, the DL solutions, the absolute error and relative error surfaces of the DL solutions, evaluated at time $t = 0.225$ for $\alpha = 0.5,1,1.5$. A detailed study of FDM numerical solution profile for \crefrange{eq:heat}{eq:heat bdry} in 2D can be found at \cite{HansenHa}.

From \cref{fig:heat_2D_range_of_good_accuracy}, we observe that $\varepsilon$-accuracy is achieved starting from time $t = 0.2$ and is preserved into some time range of future for all values of $\alpha$, in particular, the $\veps$-accuracies stay less than 0.01 for longer period of time when the value of $\alpha$ is smaller. A possible contributing factor to this behavior is that the solution of larger $\alpha$ departs further away from the initial condtions as shown in \cref{fig:FDM_alpha_0.5,,fig:FDM_alpha_1,,fig:FDM_alpha_1.5}. It is not surprising that the $\mathcal{L}^2$ relative errors for all values of $\alpha$ are strictly increasing with respect to time $t$, i.e., the closer to the training space $\Omega_T$, the more accurate the DL solution and vice versa. 
% In addition, for all values of $\alpha$ and for all times when the DL solution attaining $\veps$-accuracy ($\veps<0.01$), the average relative errors $\overline{\eps_{\text{re}}}$ defined in \cref{eq:eps_re_ave} are in general below $10^{-1.6}\approx 2.1\%$. 
The $\mathcal{L}^2$ relative error $\veps$ increases at fastest pace in time when $\alpha = 1.5$, comparing with those for $\alpha = 0.5, 1$. We also note that trapz-PiNN with the fixed NN shape $(20,5)$ is versatile enough to solve three different cases simultaneously while achieving reasonable accuracy.

\cref{fig:FDM_alpha_0.5,,fig:FDM_alpha_1,,fig:FDM_alpha_1.5} show the FDM solutions to fractional heat equations \crefrange{eq:heat}{eq:heat bdry} at time $t = 0.225$ for $\alpha = 0.5, 1, 1.5$ respectively, while \cref{fig:NN_alpha_0.5,,fig:NN_alpha_1,,fig:NN_alpha_1.5} show the corresponding DL solutions. It is hard to recognize any difference between the FDM solutions and the DL solutions from the contour plots. Maximum absolute errors of the DL solutions are of the order $10^{-3}$ for all $\alpha$. When $\alpha = 0.5$, \cref{fig:abs_err_alpha_0.5} shows the absolute error at the corners of the domain $\Omega$ are much larger than that at other regions, \cref{fig:r_err_alpha_0.5} shows that that the a majority of $\Omega$ have relative errors strictly below $10^{-2}$, in fact, $62.9\%$ of the total area of $\Omega$ have relative errors $\epsilon_{\text{re}} < 1\%$. We also note that the relative errors close to corners are on the order of $10^{-1}$ and accounts for $5.7\%$ of the area in total. In constrast, \cref{fig:abs_err_alpha_1,,fig:abs_err_alpha_1.5} show some points in the central area of $\Omega$ have the largest absolute errors for $\alpha = 1, 1.5$ respectively, while \cref{fig:r_err_alpha_1,,fig:r_err_alpha_1.5} show that these points have very low relative errors. Furthermore, $82.7\%$ and $71.1\%$ of $\Omega$ have relative errors $\epsilon_{\text{re}} < 1\%$ . Similar to the case of $\alpha = 0.5$, relative errors above $10^{-1}$ appear at region close to corners and boundaries, making up $0.3\%$ and $2.2\%$ in total area for $\alpha = 1, 1.5$ respectively. 

In general, high $\veps$-accuracy indicates small relative errors for majority of the domain $\Omega$. As can be seen from the reference solutions shown in \cref{fig:FDM_alpha_0.5,,fig:FDM_alpha_1,,fig:FDM_alpha_1.5}, the DL solutions at the regions with high relative errors have very small magnitude, compared with the interior region where the solutions have larger magnitude and small relative errors. It implies that, if true solution profile has different scales, trapz-PiNN predicts the region with moderate or large magnitude better than the region with small magnitude. 

% %%%%%%%%%%%%%%%%%%%%%%%%%%%%%%%%%%%%%%%%%%%%%%%%%%%%%%%%%%%%%%%%%%%%%%%%%%%%%%%%%%%%%%%%%%%%%% heat 2D L2 relative errors
% \begin{figure}
%     \centering
%     \includegraphics[width = 0.5\textwidth]{heat_2D/heat2D_range_of_good_accuracy.pdf}
%     \caption{The $\mathcal{L}^2$ relative errors $\veps$ (labelled as $L^2$) and average relative errors $\overline{\eps_{\text{}re}}$ (labelled as $Ave$) at time steps $0.2 + \sfrac{k}{100}, k = 0,\cdots,25$ for the DL solutions of fractional heat equations in 2D (\ref{eq:heat}) to (\ref{eq:heat bdry}) when $\alpha = 0.5, 1, 1.5$.}
%     \label{fig:heat_2D_range_of_good_accuracy}
% \end{figure}

%%%%%%%%%%%%%%%%%%%%%%%%%%%%%%%%%%%%%%%%%%%%%%%%%%%%%%%%%%%%%%%%%%%%%%%%%%%%%%%%%%%%%%%%%%%%%%% heat 2D plots

%%%%%%%%%%%%%%%%%%%%%%%%%%%%%%%%%%%%%%%%%%%%%%%%%%%%%%%%%%%%%%%%%%%%%%%%%%%%%%%%%%%%%%%%%%%%%%%%%%%%%%%%%%%%%%%%%%%%%%%%%%%%%%%%%%%%%%%%%%%%%%%%%%%%%%%%%%%%%%%%%%%%%%%%%%%%%%%%%%%%%%%%%%%%%% subsubsection heat 3D
\subsubsection{3D case} \label{subsubsection:heat 3D}
\begin{figure}[h]
    \begin{subfigure}[t]{0.24\textwidth}
         \includegraphics[width=\textwidth]{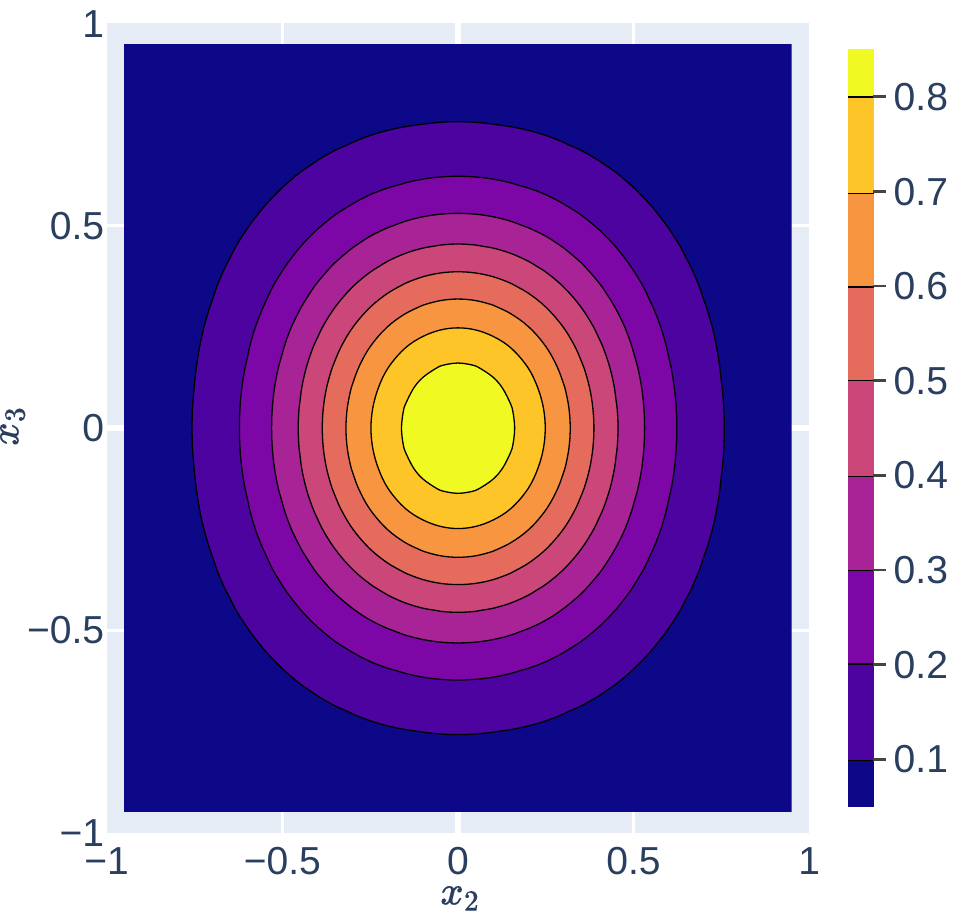}
         \caption{FDM solution \\ $x_1=-0.1$}
         \label{fig:FDM_mid_heat_3D}
     \end{subfigure}
     \hfill
        \begin{subfigure}[t]{0.24\textwidth}
         \includegraphics[width=\textwidth]{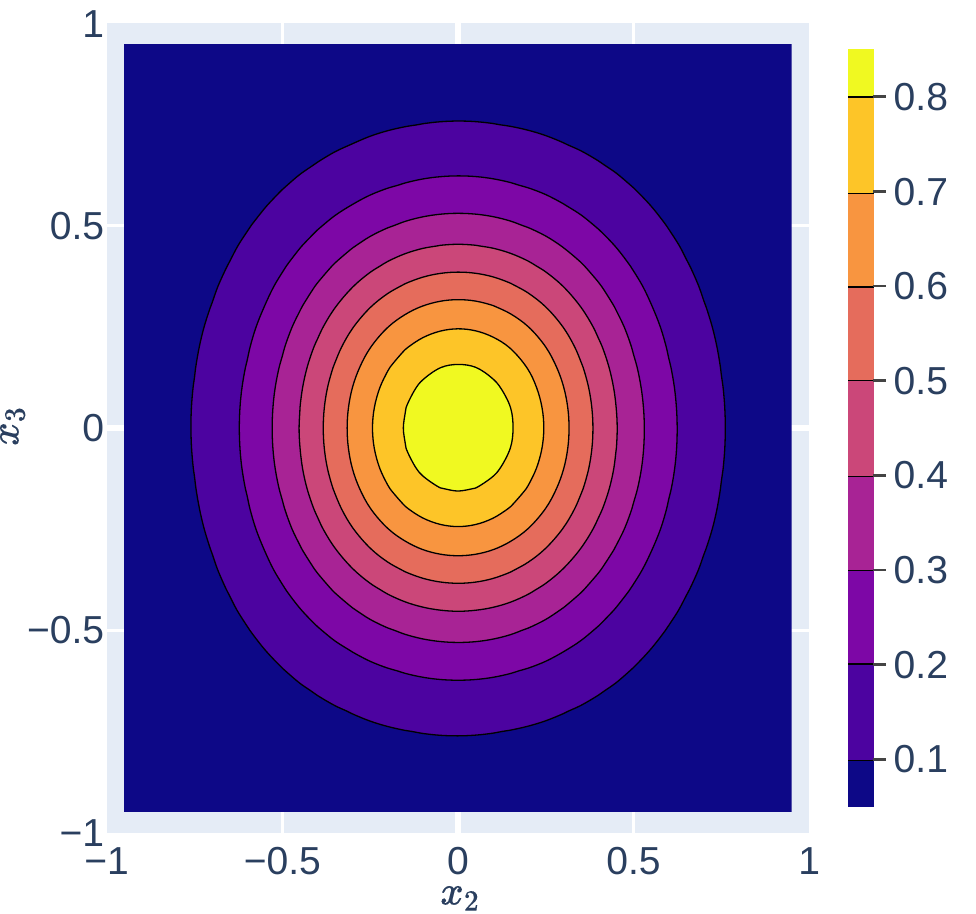}
         \caption{DL solution \\ $x_1=-0.1$}
         \label{fig:NN_mid_heat_3D}
     \end{subfigure}
      \hfill
         \begin{subfigure}[t]{0.24\textwidth}
         \includegraphics[width=\textwidth]{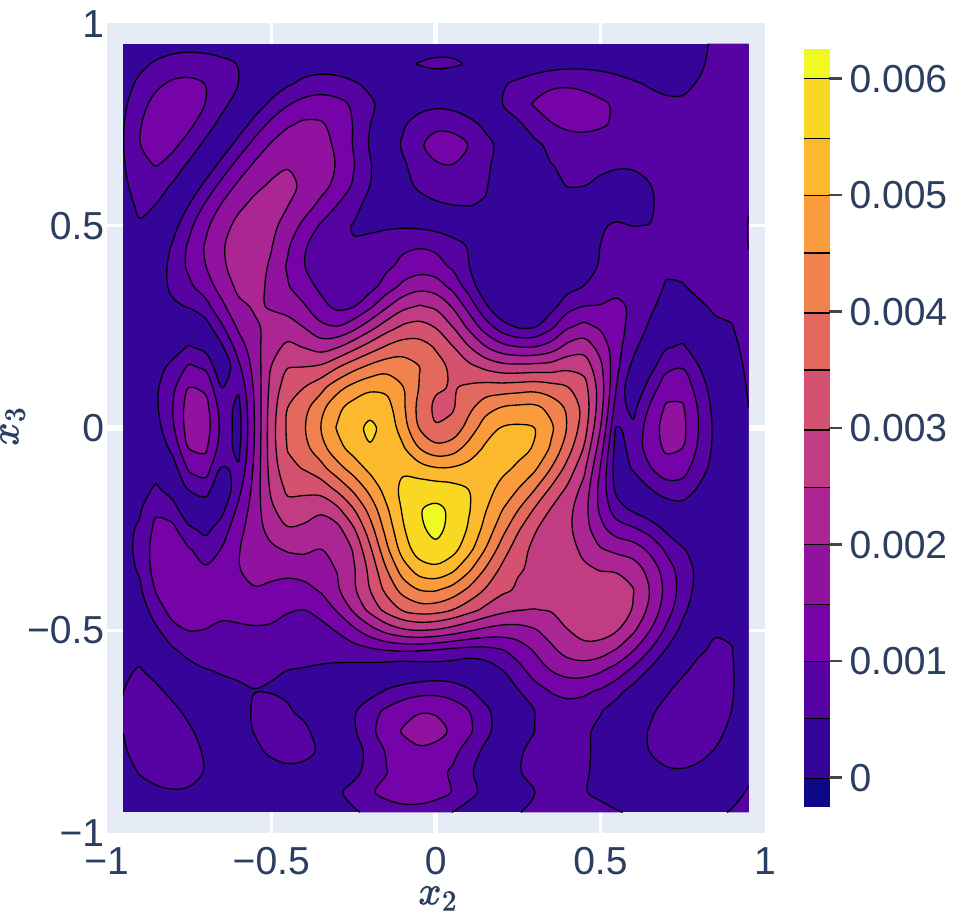}
         \caption{$\eps_{\text{abs}}$ for $x_1=-0.1$}
         \label{fig:abs_err_mid_heat_3D}
     \end{subfigure}
        \hfill
         \begin{subfigure}[t]{0.24\textwidth}
         \includegraphics[width=\textwidth]{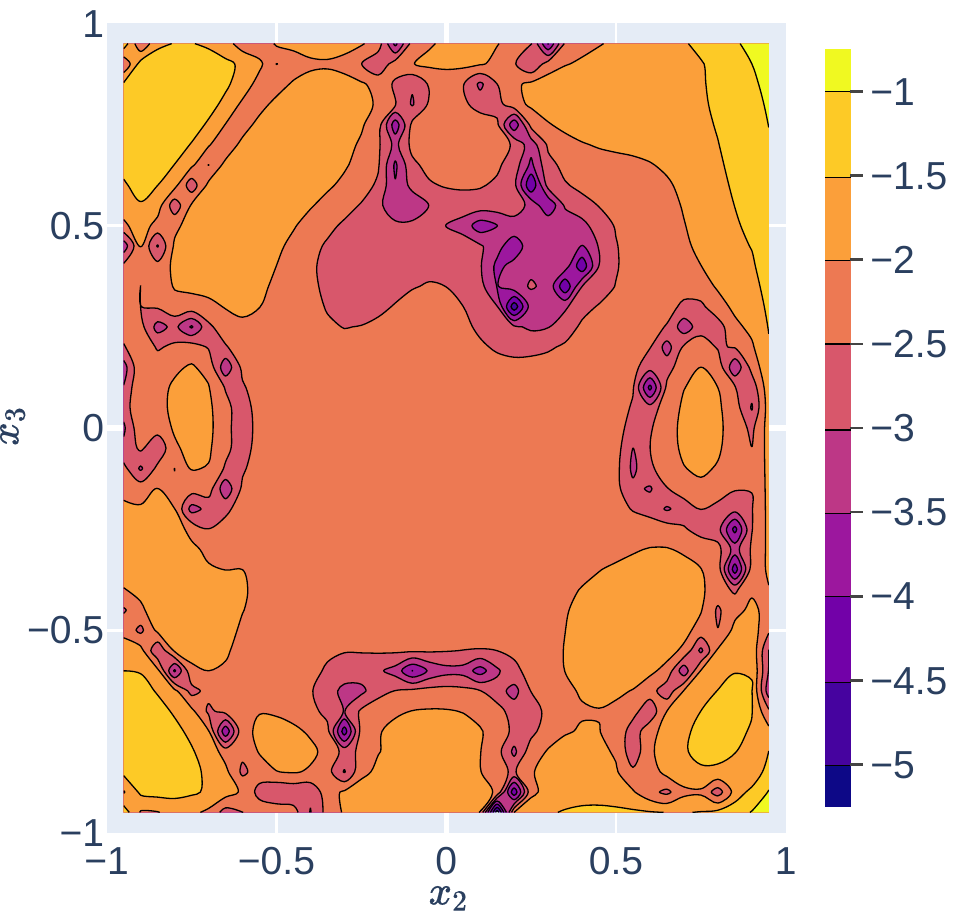}
         \caption{$\log_{10}(\epsilon_{\text{re}})$ for $x_1=-0.1$ }
         \label{fig:r_err_mid_heat_3D}
     \end{subfigure}
     \hfill
    \begin{subfigure}[t]{0.24\textwidth}
         \includegraphics[width=\textwidth]{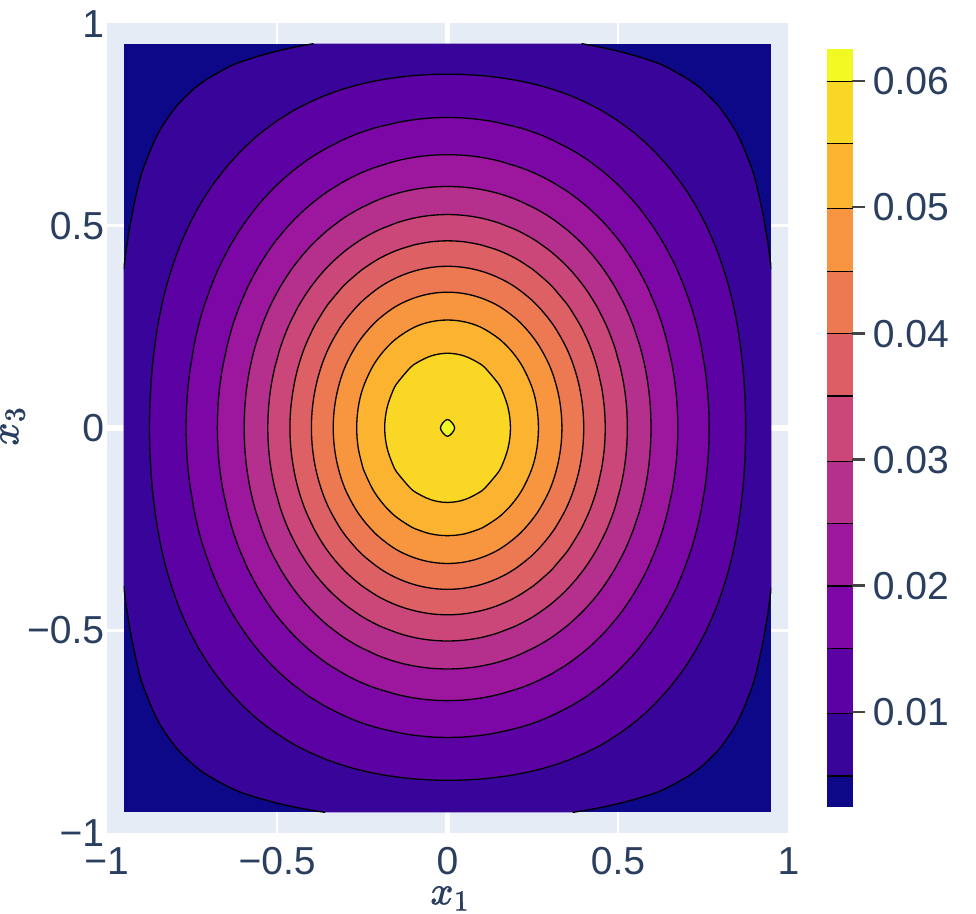}
         \caption{FDM solution \\ $x_2=0.85$}
         \label{fig:FDM_top_heat_3D}
     \end{subfigure}
     \hfill
     \begin{subfigure}[t]{0.24\textwidth}
         \includegraphics[width=\textwidth]{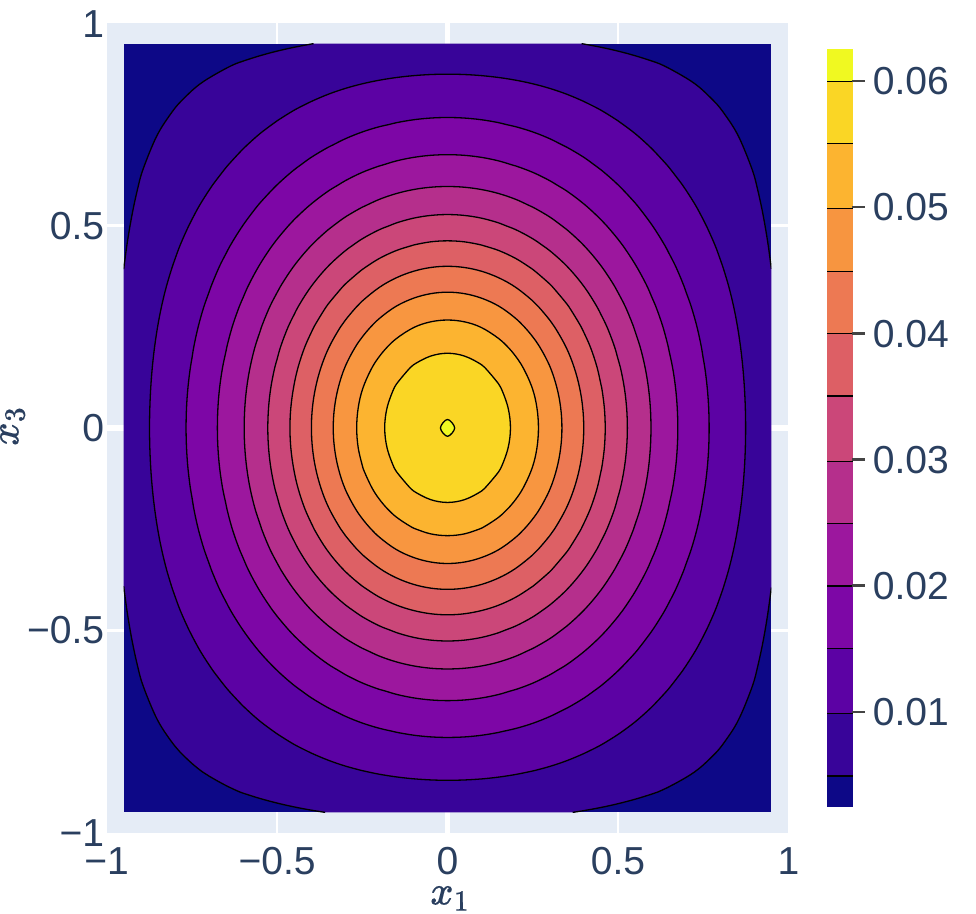}
         \caption{DL solution \\ $x_2=0.85$}
         \label{fig:NN_top_heat_3D}
     \end{subfigure}
     \hfill
     \begin{subfigure}[t]{0.24\textwidth}
         \centering
         \includegraphics[width=\textwidth]{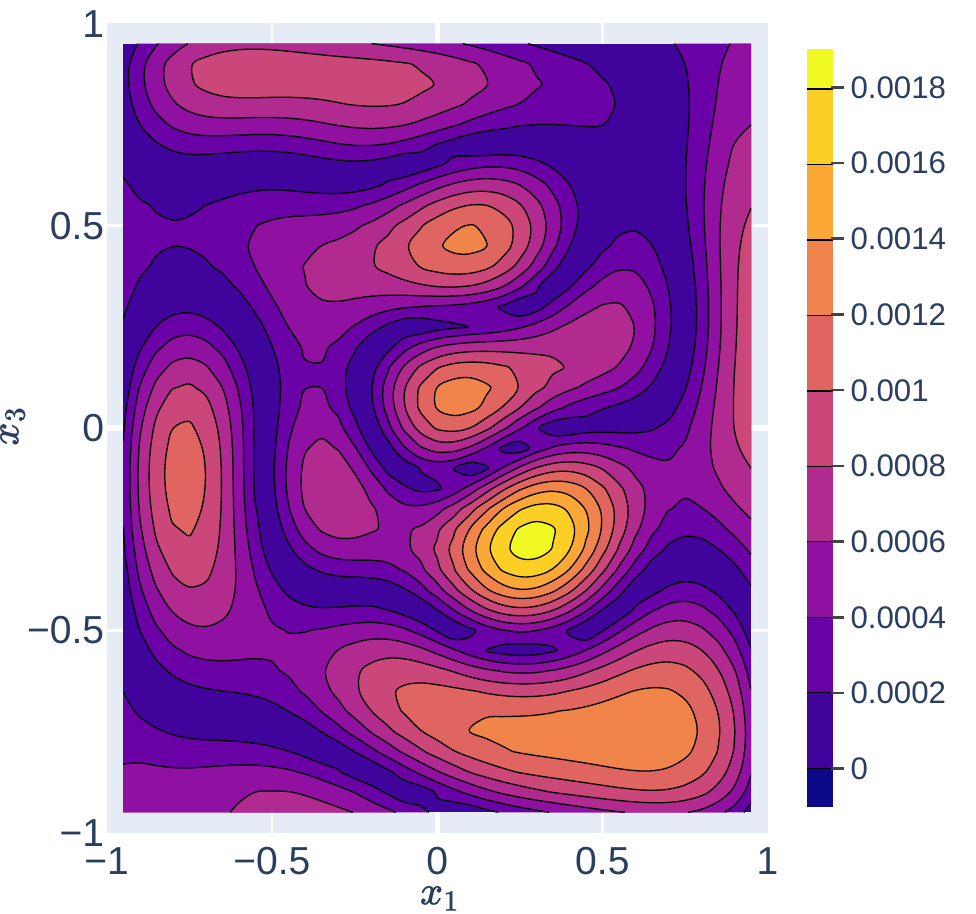}
         \caption{$\eps_{\text{abs}}$ for $x_2=0.85$}
         \label{fig:abs_err_top_heat_3D}
     \end{subfigure}
    \hfill
         \begin{subfigure}[t]{0.24\textwidth}
         \includegraphics[width=\textwidth]{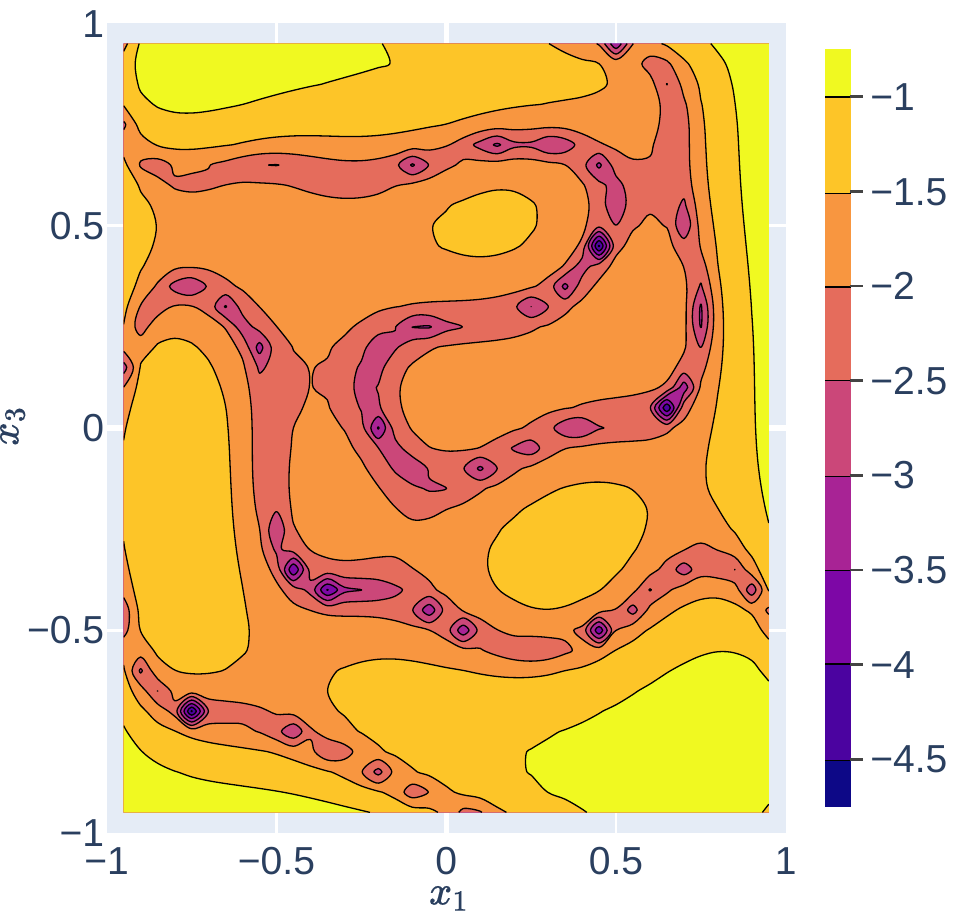}
         \caption{$\log_{10}(\epsilon_{\text{re}})$ for $x_2=0.85$ }
         \label{fig:r_err_top_heat_3D}
     \end{subfigure}
    \caption{The contour plots of the FDM solutions, the DL solutions, the absolute and relative errors of the cross-sections $x_1 = -0.1$ and $x_2 = 0.85$ of the DL solutions for the fractional heat equations in 3D \crefrange{eq:heat}{eq:heat bdry} at $t = 0.21$ when $\alpha = 1$. }
    \label{fig:Heat_3D}
\end{figure}

In 3D, the hyperparameters for trapz-PiNN are chosen as the NN shape $(20, 7)$, the learning rate $0.001$, the epoch $3\times 10^5$, the batch size $100$, the space resolution $\sfrac{1}{20}$ and the time resolution $0.01$. The $\mathcal{L}^2$ relative errors $\veps$ at $t = 0.2 + \sfrac{k}{100}, k = 0,\cdots,5$ are recorded at \cref{tab:L2_r_err_heat_3D}. \cref{fig:Heat_3D} presents the contour plots of the FDM and DL solutions, the absolute and relative errors of the DL solutions evaluated at $t = 0.21$ for the cross-sections $x_1 = -0.1$ and $x_2 = 0.85$, respecitvely. Due to the limitation of RAM, we use moderate space resolution $\frac{1}{20}$ for both FDM and DL solutions. 

From \cref{tab:L2_r_err_heat_3D}, we observe that the $\varepsilon$-accuracy ($\veps < 0.01$) is achieved starting from $t = 0.2$ and preserved before $t = 0.25$. Similar to the 2D case, the FDM solutions shown in \cref{fig:FDM_mid_heat_3D,,fig:FDM_top_heat_3D} and the corresponding DL solutions shown in \cref{fig:NN_mid_heat_3D,,fig:NN_top_heat_3D} are indistinguishable. \cref{fig:abs_err_mid_heat_3D} shows the maximum absolute error $\eps_{\text{abs}}$ attained at some central points in the cross-section $x_1 = -0.1$. \cref{fig:r_err_mid_heat_3D} shows that the majority of the cross-section have relative error below or around $10^{-2}$. \cref{fig:abs_err_top_heat_3D} also shows the absolute error peaks at some central points in the cross-section $x_2 = 0.85$. \cref{fig:r_err_top_heat_3D} shows the pointwise relative error $\eps_{\text{re}}$ at a significant part of the area close to the boundary of the cross-section $x_2 = 0.85$ is on the order of $10^{-1}$ while the other areas is below $10^{-1.5} \approx 3.2\%$. Furthermore, $76.1\%$ of $\Omega$ have the relative error $\epsilon_{\text{re}} < 3\%$ while the regions with relative error $\eps_{\text{re}} > 10\%$ occupy $5.2\%$ in $\Omega$.

We see that moderate space resolution $h$ reduces the size of training data $\mathcal{T}$ and therefore the range of $\varepsilon$-accuracy, compared with the 2D counter-part. From \cref{fig:r_err_mid_heat_3D,,fig:r_err_top_heat_3D}, we find that the cross-section $x_1 = -0.1$ has better overall accuracy than that of $x_2 = 0.85$, possibly since the latter one is closer to boundary of the cube $\Omega$ in $\R^3$. We also know from the FDM solutions shown in \cref{fig:FDM_mid_heat_3D,,fig:FDM_top_heat_3D} that the magnitude of solution profile for the cross-section $x_2 = 0.85$ is small. It follows again that trapz-PiNN predicts better in the region with moderate or large magnitude than region with small magnitude.

%%%%%%%%%%%%%%%%%%%%%%%%%%%%%%%%%%%%%%%%%%%%%%%%%%%%%%%%%%%%%%%%%%%%%%%%%%%%%%%%%%%%%%%%%%%%%%% heat 3D L2 r errors
\begin{table}[h]
    \centering
     \caption{The $\mathcal{L}^2$ relative errors $\veps$ at time steps $0.2 + \sfrac{k}{100}, k = 0,\cdots,5$ for the DL solutions of fractional heat equations in 3D (\crefrange{eq:heat}{eq:heat bdry}) with $\alpha = 1$}
    \begin{tabular}{c|cccccc}
    \hline\hline
    $T_{\text{pred}}$ & $t = 0.2$ & $t = 0.21$ & $t = 0.22$ & $t = 0.23$ & $t = 0.24$ & $t = 0.25$ \\
    \hline
    $\mathcal{L}^2$ relative errors  & 6.856e-3   & 7.311e-3  & 7.838e-3 & 8.458e-3 & 9.196e-3  & 1.001e-2  \\
    % Average relative errors & 0.0254 & 0.0253 & 0.0255 & 0.026 & 0.027 & 0.028 \\
      \hline\hline
    \end{tabular}
    \label{tab:L2_r_err_heat_3D}
\end{table}

%%%%%%%%%%%%%%%%%%%%%%%%%%%%%%%%%%%%%%%%%%%%%%%%%%%%%%%%%%%%%%%%%%%%%%%%%%%%%%%%%%%%%%%%%%%%%%%%% heat_3D plots

%%%%%%%%%%%%%%%%%%%%%%%%%%%%%%%%%%%%%%%%%%%%%%%%%%%%%%%%%%%%%%%%%%%%%%%%%%%%%%%%%%%%%%%%%%%%%%%%%%%%%%%%%%%%%%%%%%%%%%%%%%%%%%%%%%%%%%%%%%%%%%%%%%%%%%%%%%%%%%
\subsection{FPEs with Ornstein–Uhlenbeck Potential in 2D} \label{subsection:OU}
\begin{figure}[h]
    \begin{subfigure}[b]{0.24\textwidth}
         \centering
         \includegraphics[width=\textwidth]{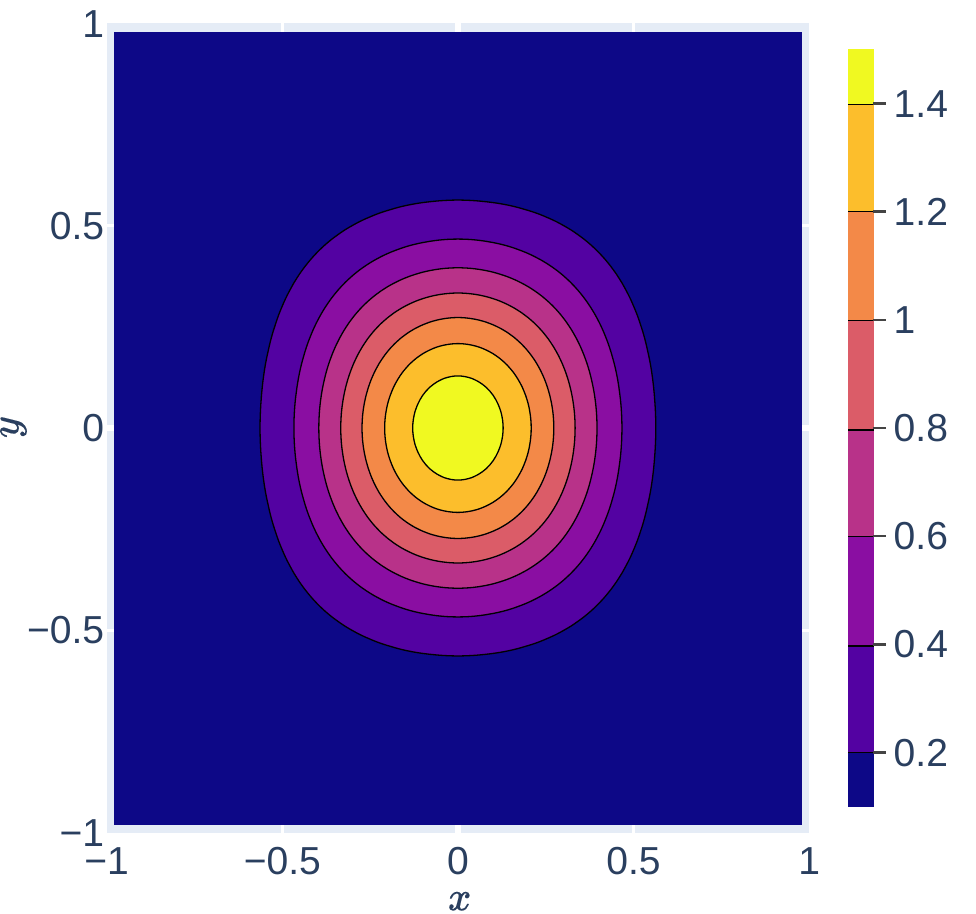}
         \caption{FDM solution $\alpha = 0.5$}
         \label{fig:FDM_alpha_0.5_OU}
     \end{subfigure}
     \hfill
         \begin{subfigure}[b]{0.24\textwidth}
         \centering
         \includegraphics[width=\textwidth]{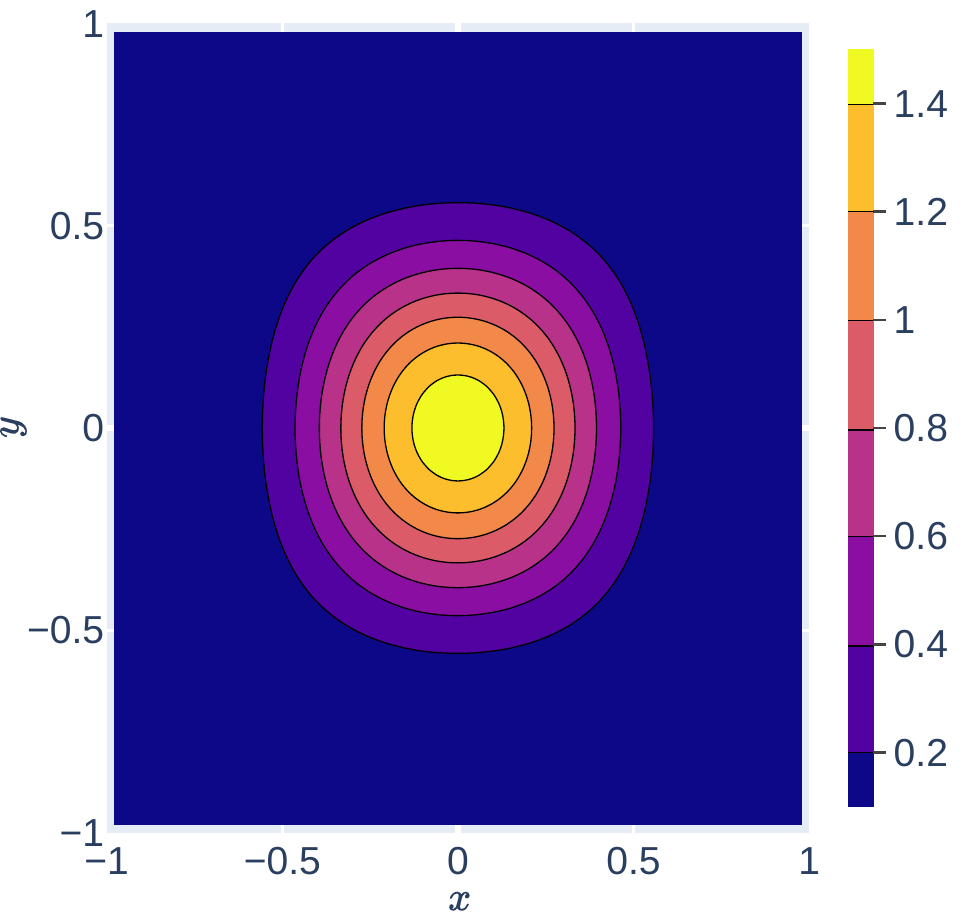}
         \caption{DL solution $\alpha = 0.5$}
         \label{fig:NN_alpha_0.5_OU}
     \end{subfigure}
      \hfill
         \begin{subfigure}[b]{0.24\textwidth}
         \centering
         \includegraphics[width=\textwidth]{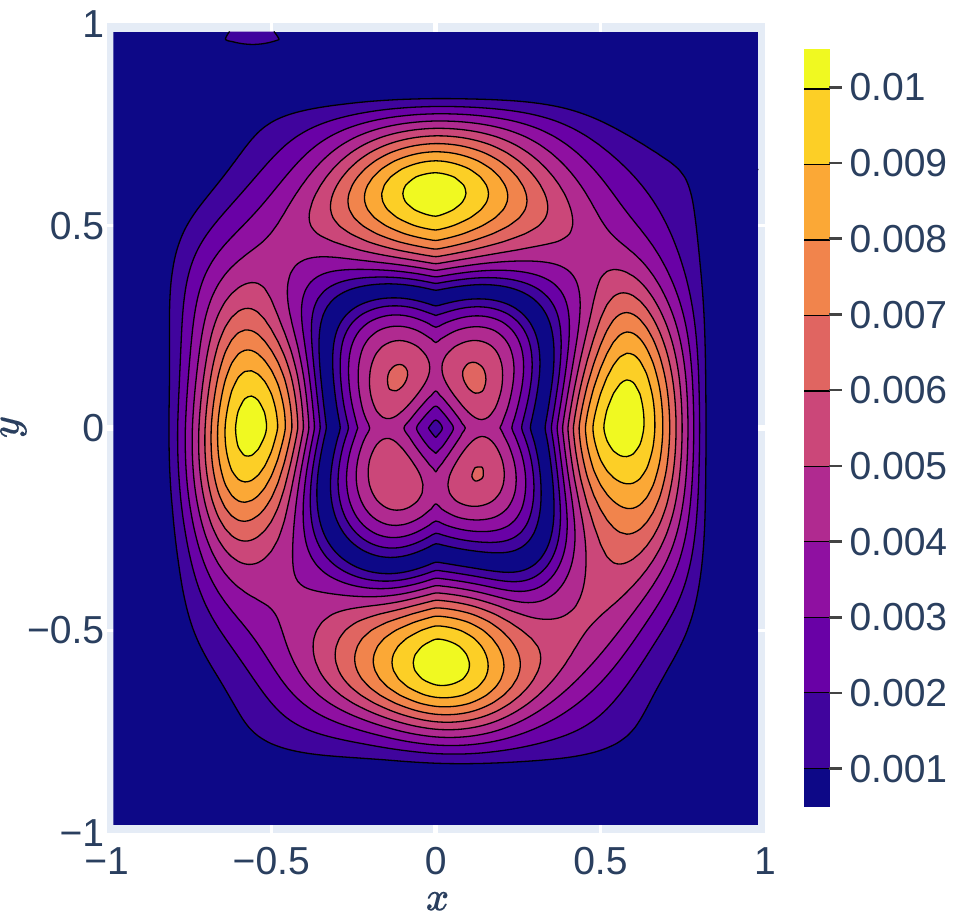}
         \caption{$\epsilon_{\text{abs}}$ for $\alpha = 0.5$}
         \label{fig:abs_err_alpha_0.5_OU}
     \end{subfigure}
        \hfill
         \begin{subfigure}[b]{0.24\textwidth}
         \centering
         \includegraphics[width=\textwidth]{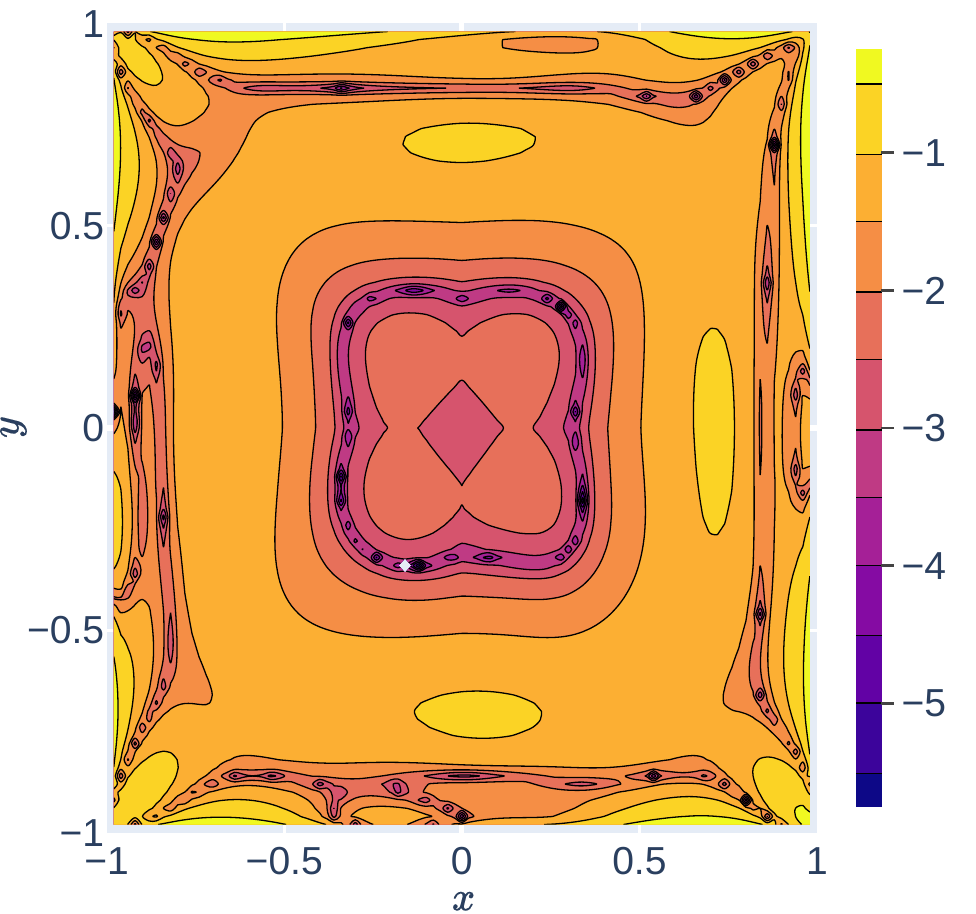}
         \caption{$\log_{10}(\epsilon_{\text{re}})$ for $\alpha = 0.5$ }
         \label{fig:r_err_alpha_0.5_OU}
     \end{subfigure}
     \hfill
         \begin{subfigure}[b]{0.24\textwidth}
         \centering
         \includegraphics[width=\textwidth]{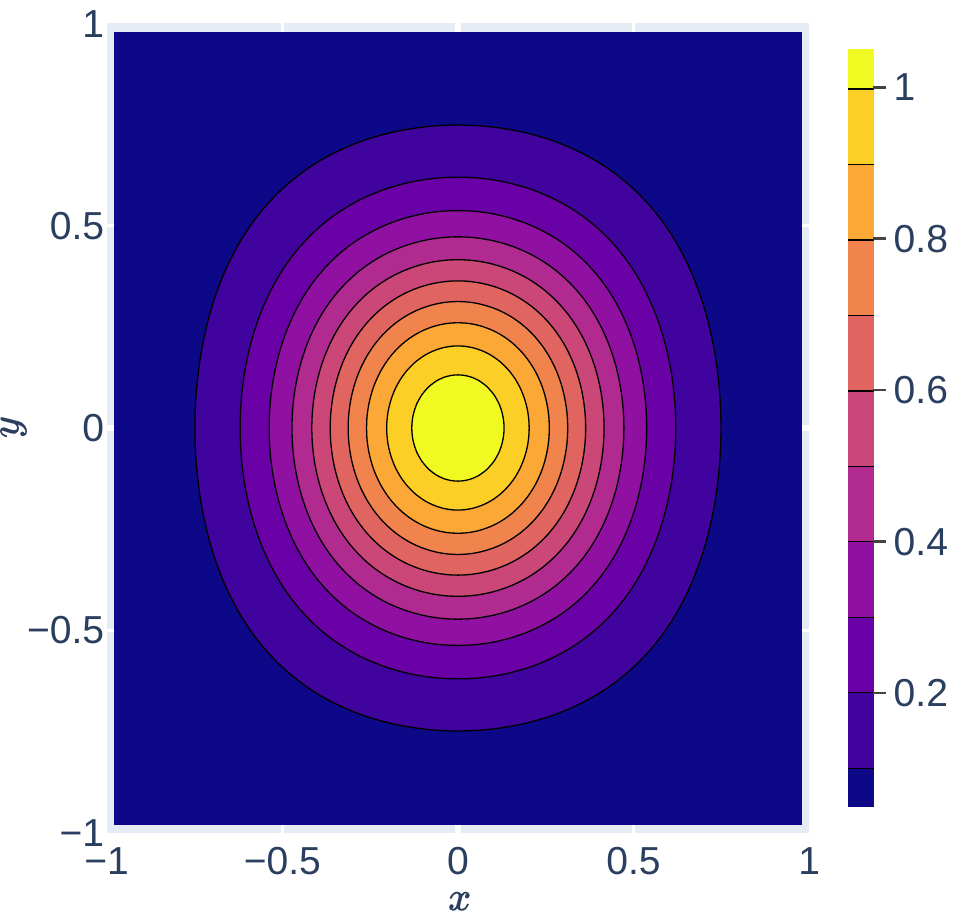}
         \caption{FDM solution $\alpha = 1$ }
         \label{fig:FDM_alpha_1_OU}
     \end{subfigure}
        \hfill
         \begin{subfigure}[b]{0.24\textwidth}
         \centering
         \includegraphics[width=\textwidth]{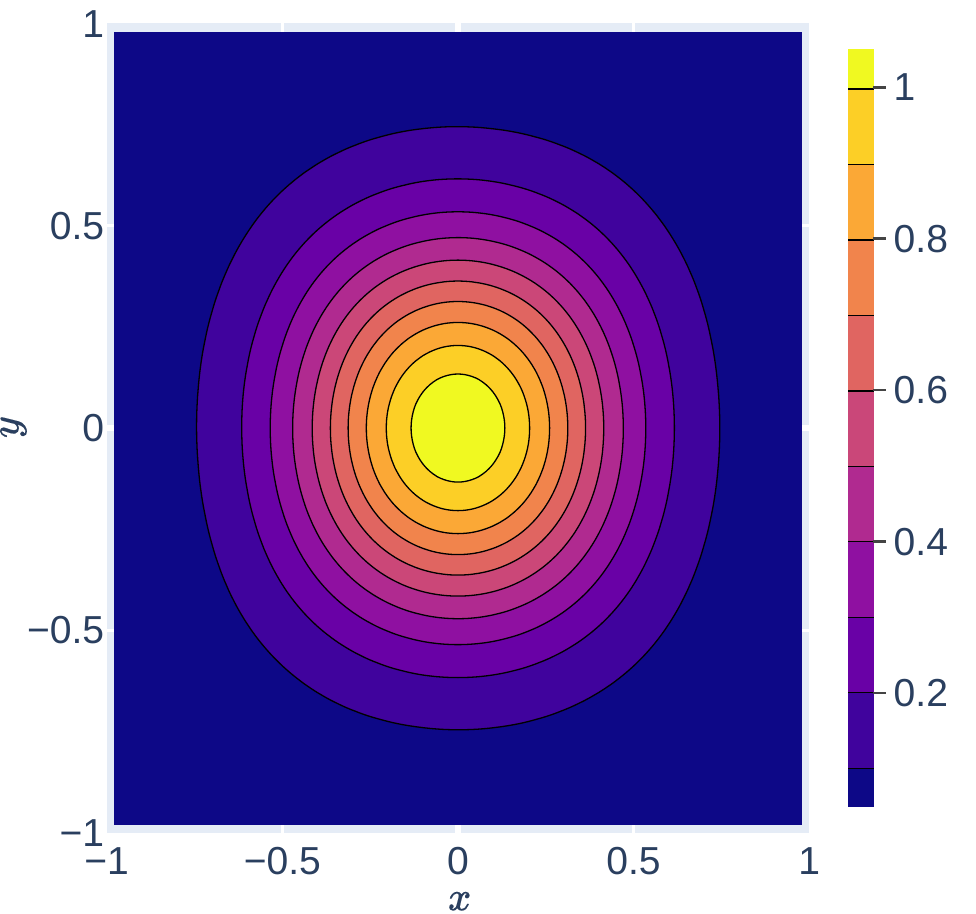}
         \caption{DL solution $\alpha = 1$ }
         \label{fig:NN_alpha_1_OU}
     \end{subfigure}
             \hfill
         \begin{subfigure}[b]{0.24\textwidth}
         \centering
         \includegraphics[width=\textwidth]{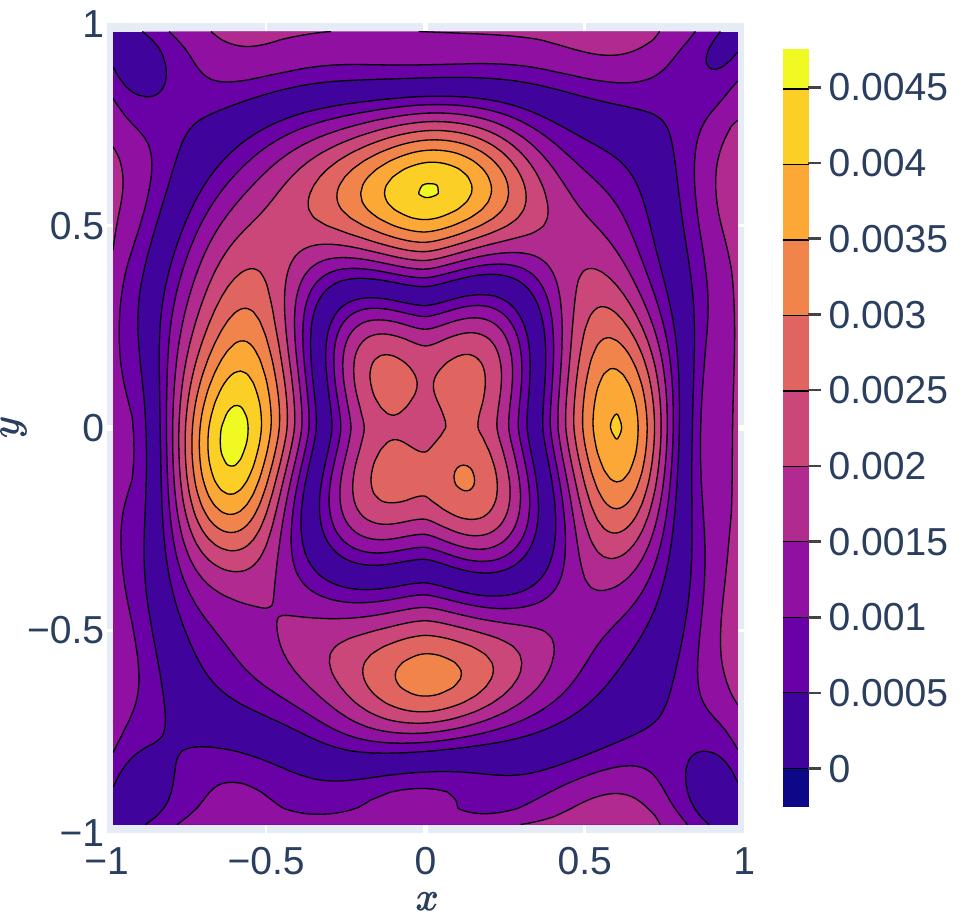}
         \caption{$\epsilon_{\text{abs}}$ for $\alpha = 1$ }
         \label{fig:abs_err_alpha_1_OU}
     \end{subfigure}
             \hfill
         \begin{subfigure}[b]{0.24\textwidth}
         \centering
         \includegraphics[width=\textwidth]{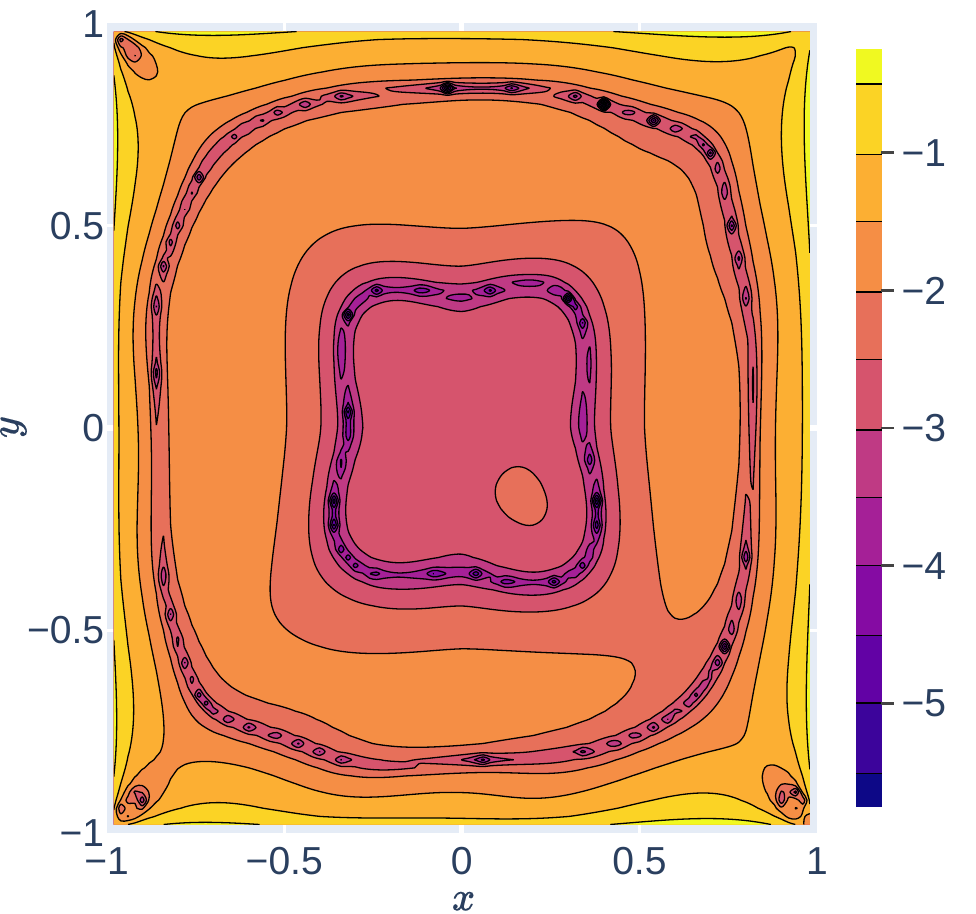}
         \caption{$\log_{10}(\epsilon_{\text{re}})$ for $\alpha = 1$ }
         \label{fig:r_err_alpha_1_OU}
     \end{subfigure}
          \hfill
         \begin{subfigure}[b]{0.24\textwidth}
         \centering
         \includegraphics[width=\textwidth]{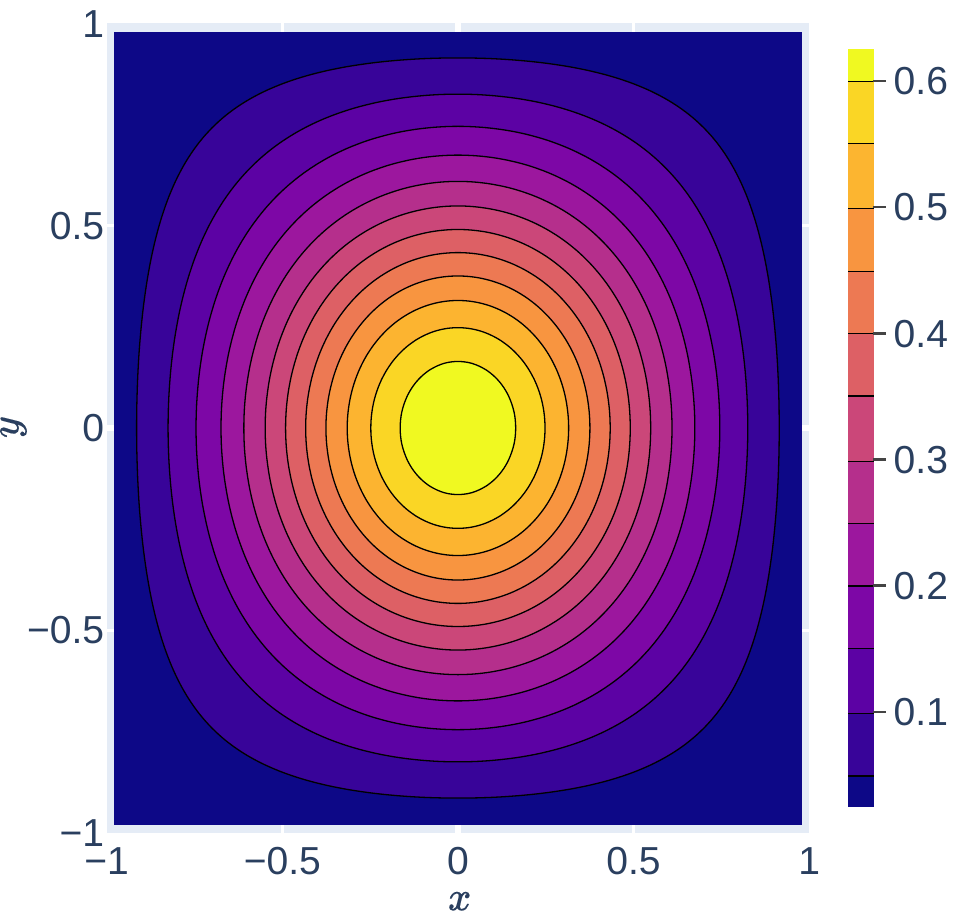}
         \caption{FDM solution $\alpha = 1.5$ }
         \label{fig:FDM_alpha_1.5_OU}
     \end{subfigure}
        \hfill
         \begin{subfigure}[b]{0.24\textwidth}
         \centering
         \includegraphics[width=\textwidth]{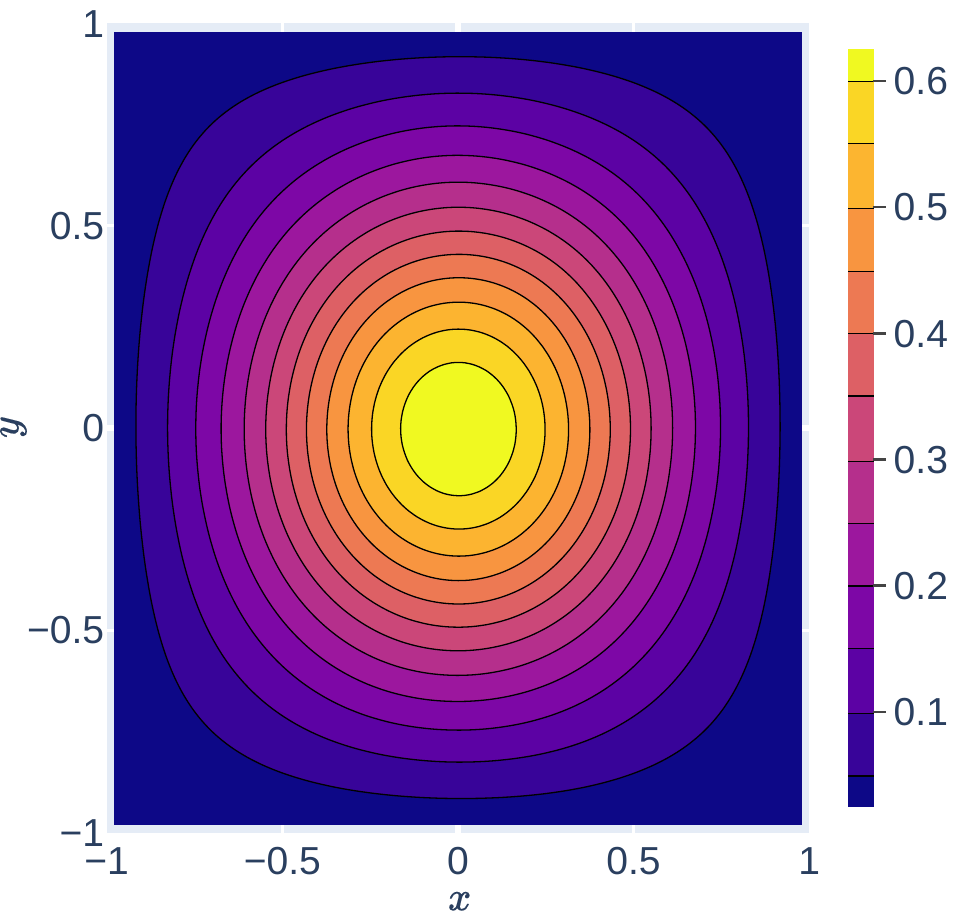}
         \caption{DL solution $\alpha = 1.5$ }
         \label{fig:NN_alpha_1.5_OU}
     \end{subfigure}
             \hfill
         \begin{subfigure}[b]{0.24\textwidth}
         \centering
         \includegraphics[width=\textwidth]{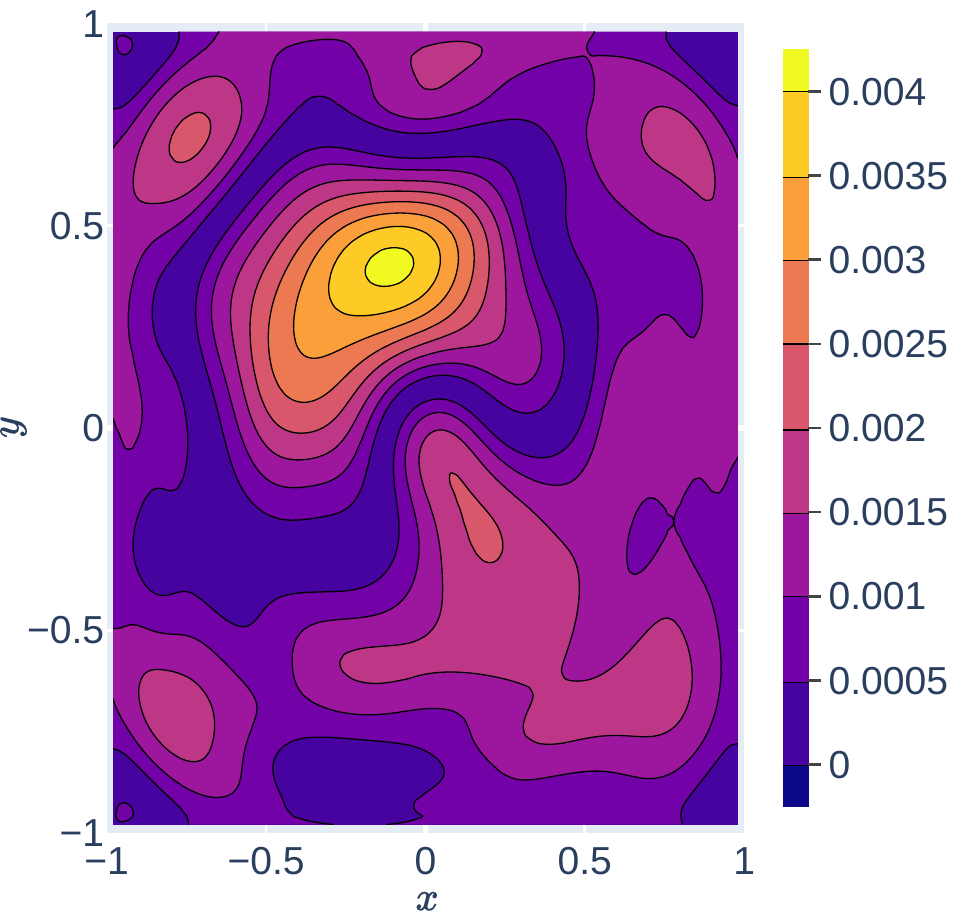}
         \caption{$\epsilon_{\text{abs}}$ for $\alpha = 1.5$ }
         \label{fig:abs_err_alpha_1.5_OU}
     \end{subfigure}
             \hfill
         \begin{subfigure}[b]{0.24\textwidth}
         \centering
         \includegraphics[width=\textwidth]{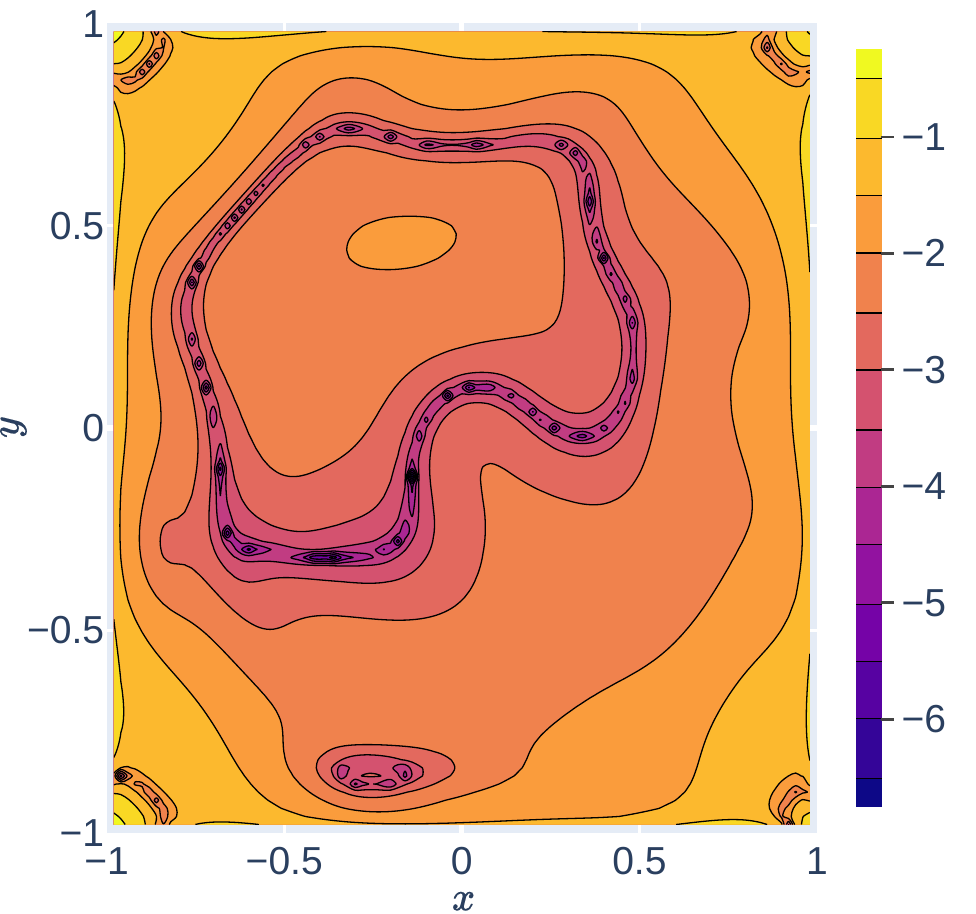}
         \caption{$\log_{10}(\epsilon_{\text{re}})$ for $\alpha = 1.5$ }
         \label{fig:r_err_alpha_1.5_OU}
     \end{subfigure}
    \caption{The contour plots of the FDM solutions, the DL solutions, the absolute and relative errors of DL solutions for FPE with O-U potential in 2D \crefrange{eq:OU}{eq:OU bdry} sampled at $t = 0.2$ for $\alpha = 0.5, 1, 1.5$. }
    \label{fig:OU}
\end{figure}

We consider the FPE with O-U potential in 2D: \begin{align}
    \partial_tu(x,t) &= -(\partial_1(x_1u(x,t))-\partial_2(x_2u(x,t))) -(-\Delta)^{\frac{\alpha}{2}}u(x,t), \label{eq:OU}\\
    u(x,0) &= \left(\frac{315}{216}\right)^2((1-x_1^2)(1-x_2^2))_+^4, \label{eq:OU init cond}\\
    u(x,t) &= 0,\quad x\not \in (-1,1)^2, \label{eq:OU bdry}
\end{align} with $\alpha = 0.5, 1, 1.5$. The solution $u$ to the Fokker-Planck equation \crefrange{eq:OU}{eq:OU bdry} is the probability density function corresponding to the stochastic differential equations in 2D\begin{align}
    \d X_t = -X_t\d t + \d L^{\alpha}_t,\quad X_0 = X,
\end{align} where $L_t^{\alpha}$ is the $\alpha$-stable L\'evy process and random variable $X$ has distribution as $u(x,0)$, with stochastic process vanishes outside domain $\Omega$. Without the $\alpha$-stable noise, the determinstic dynamical system drives the process to the unique stable point, the origin $(0,0)$. Hyperparameters for trapz-PiNNs are chosen as the NN shape $(20, 6)$, the learning rate $10^{-3}$, the epoch $2\times 10^5$, the batch size $64$, the space resolution $\sfrac{1}{50}$ and the time resolution $5\times 10^{-3}$. \cref{fig:OU_range_of_good_accuracy} records the $\mathcal{L}^2$ relative errors $\veps$ at $t = 0.2 + \sfrac{k}{100}, k = 0,\cdots, 10$. \cref{fig:OU} presents the FDM and DL solutions and the absolute and relative error surfaces of the DL solutions evaluated at $t = 0.2$ for $\alpha = 0.5,1,1.5$ respecitvely. For more information on FDM numerical solution profile for equations \crefrange{eq:OU}{eq:OU bdry} in 2D, we refer to \cite{HansenHa}.

From \cref{fig:OU_range_of_good_accuracy}, we observe that the $\varepsilon$-accuracy ($\veps<0.01$) is achieved only at time $t = 0.2$ for $\alpha = 0.5$ under the effect of the O-U potential, on the other hand, the $\veps$-accuracy is preserved up to time $t = 0.25$ for $\alpha = 1.5$ and up to $t = 0.3$ for $\alpha = 1$. 
% Average relative error $\overline{\eps_{\text{re}}}$ for $\alpha = 0.5$ at $t = 0.2$ is around $5\%$, whereas for $\alpha = 1$, $\overline{\eps_{\text{re}}}$ starts from $3\%$ at $t = 0.2$ and finishes around $5\%$ at $t = 0.3$; For $\alpha = 1.5$, $\overline{\eps_{\text{re}}}$ starts from $2\%$ at $t = 0.2$ and finishes around $3\%$ at $t = 0.24$. 
$\mathcal{L}^2$ relative errors are increasing with respect to time $t$ for all values of $\alpha$. The rate of increase is highest for $\alpha = 1.5$.

\cref{fig:FDM_alpha_0.5_OU,,fig:FDM_alpha_1_OU,,fig:FDM_alpha_1.5_OU} show the FDM solutions to Fokker-Planck equation with O-U potential in 2D \crefrange{eq:OU}{eq:OU bdry} at time $t = 0.2$ for $\alpha = 0.5, 1, 1.5$ respectively, while \cref{fig:NN_alpha_0.5_OU,,fig:NN_alpha_1_OU,,fig:NN_alpha_1.5_OU} are the corresponding DL solutions. They are hard to tell apart. When $\alpha = 0.5$, \cref{fig:abs_err_alpha_0.5_OU} shows maximum absolute error is of the order $10^{-2}$ and is attained at some interior points in the domain, \cref{fig:r_err_alpha_0.5_OU} shows these peak points have the relative errors $\eps_{\text{re}}$ around $ 10^{-1.25}\approx 5.6\%$. Together with the FDM solution profile shown in \cref{fig:FDM_alpha_0.5_OU}, we see that the area with the relative errors $\eps_{\text{re}}$ greater than $3\%$ concertrates in the area with small magnitude (deep blue area) of the solution. When $\alpha = 0.5, 1$, \cref{fig:abs_err_alpha_1_OU,,fig:abs_err_alpha_1.5_OU} show the maximum absolute errors are below $5\times 10^{-3}$. Together with the FDM solution profiles shown in \cref{fig:FDM_alpha_1_OU,,fig:FDM_alpha_1.5_OU}, \cref{fig:r_err_alpha_1_OU,,fig:r_err_alpha_1.5_OU} show the points attaining maximum absolute error have relative error below $10^{-1.5}\approx 3.2\%$. Furthurmore, for $\alpha = 1 \text{ and } 1.5$, more than $76.9\%$ and $80\%$ of the total area has the relative error below $3\%$ respectively and the area with large relative errors locates at the area with the solution of small magnitude. They account for $8.6\%$ and $2.1\%$ of the total area, respectively.

Under the effect of O-U potential, $\veps$-accuracy and range of $\veps$-accuracy are no longer similar to the case without O-U potential in \cref{subsubsection:heat 2D}. O-U potential drives the solution profile more higher at the origin and for $\alpha < 1$, it drives the profile near the boundary of the domain close to zero \cite{HansenHa}. This may explain the trapz-PiNN can predict accurately the solution for $\alpha = 0.5$ at $t = 0.2$ only but gives much better predictions for the cases of $\alpha = 1$ and $1.5$.

% With finer space resolution $\sfrac{1}{50}$ and more epoch $6\times 10^5$, we found that trapz-PiNN is able to predicts accurately up to $t = 0.21$.

%%%%%%%%%%%%%%%%%%%%%%%%%%%%%%%%%%%%%%%%%%%%%%%%%%%%%%%%%%%%%%%%%%%%%%%%%%%%%%%%%%%%%%%%%%%%%%% OU L2 relative errors
% \begin{figure}[ht]
%     \centering
%     \includegraphics[width = 0.5\textwidth]{OU/OU_range_of_good_accuracy.pdf}
%     \caption{$\mathcal{L}^2$ relative errors ($L^2$) and average relative errors ($Ave$) at time steps $0.2 + \sfrac{k}{100}, k = 0,\cdots,10$ for DL solutions of FPEs with O-U potential in 2D \crefrange{eq:OU}{eq:OU init cond} when $\alpha = 0.5, 1, 1.5$.}
%     \label{fig:OU_range_of_good_accuracy}
% \end{figure}
%%%%%%%%%%%%%%%%%%%%%%%%%%%%%%%%%%%%%%%%%%%%%%%%%%%%%%%%%%%%%%%%%%%%%%%%%%%%%%%%%%%%%%%%%% OU plots

%%%%%%%%%%%%%%%%%%%%%%%%%%%%%%%%%%%%%%%%%%%%%%%%%%%%%%%%%%%%%%%%%%%%%%%%%%%%%%%%%%%%%%%%%%%%%%%%%%%%%%%%%%%%%%%%%%%%%%%%%%%%%%%%%%%%%%%%%%%%%%%%%%%%%%%%%%%%%%%%%%%%%%%%%%%%%%%%%%%%%%%%%%%%%%%%%%%%%%%%%%%%%%%%%%%%%%%%%%%%%%%%%%%%%%%%%%%%%%%%%%%%%%%%%%%%%%%%%%%%%%%%%%%%
\subsection{A modified loss function}\label{subsection:mod loss}
From the numerical results in \cref{subsection:frac_heat_2D,,subsection:OU}, we find that the trapz-PiNN is able to predict solutions with good overall accuracy and more accurate in the region with large or moderate magnitude than the region with small magnitude, when the true solution profile has different scales. The loss function MSE defined by \cref{eq:loss_function} is a metric of global accuracy, rather than a pointwise one. If physical observations of solutions or high-fidelity simulated data is available, we propose an effective loss function to address above-mentioned issue. 

\begin{figure}[h]
     \begin{subfigure}[b]{0.24\textwidth}
         \centering
         \includegraphics[width=\textwidth]{OU/fig_max_err_T_0.2_alpha_0.5_eps_1_2022-05-27.pdf}
         \caption{$\epsilon_{\text{abs}}$ - original}
         \label{fig:abs_err_alpha_0.5_mod_loss_original}
     \end{subfigure}
     \hfill
     \begin{subfigure}[b]{0.24\textwidth}
         \centering
         \includegraphics[width=\textwidth]{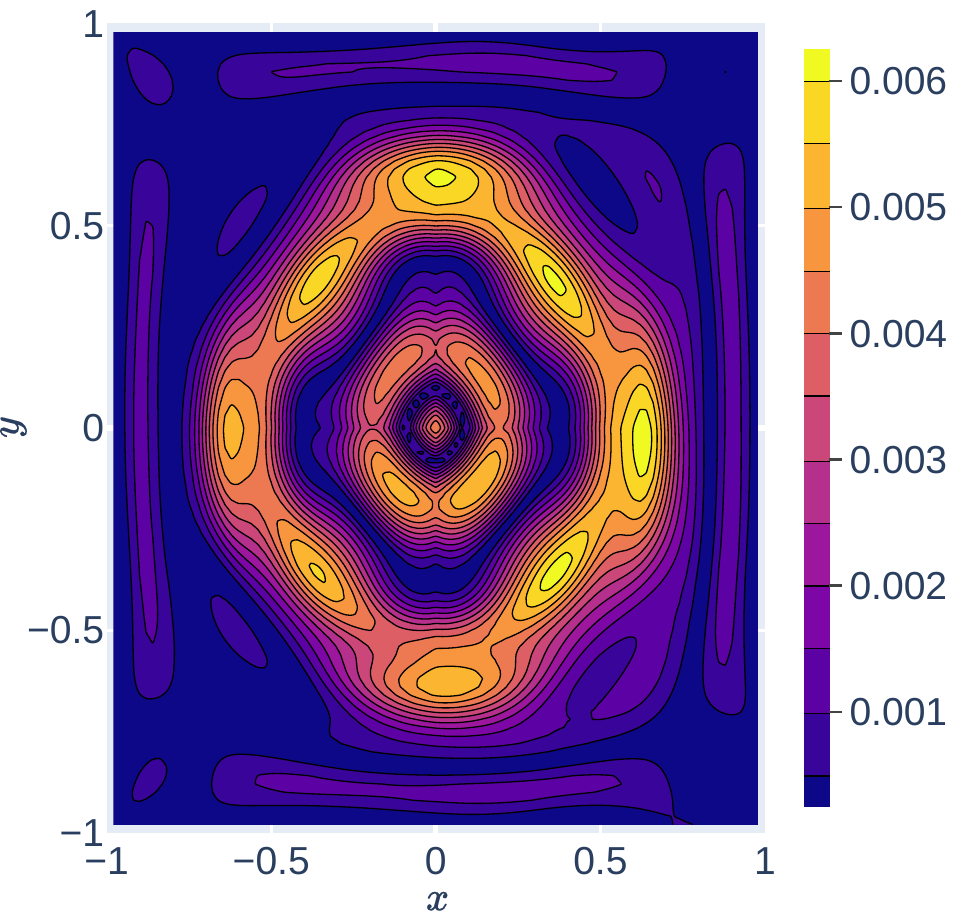}
         \caption{$\epsilon_{\text{abs}}$ - new }
         \label{fig:abs_err_alpha_0.5_mod_loss_new}
     \end{subfigure}
         \hfill
         \begin{subfigure}[b]{0.24\textwidth}
         \centering
         \includegraphics[width=\textwidth]{OU/fig_r_err_T_0.2_alpha_0.5_eps_1_2022-05-27.pdf}
         \caption{$\log_{10}(\epsilon_{\text{re}})$ - original }
         \label{fig:r_err_alpha_0.5_mod_loss_original}
     \end{subfigure}
         \hfill
         \begin{subfigure}[b]{0.24\textwidth}
         \centering
         \includegraphics[width=\textwidth]{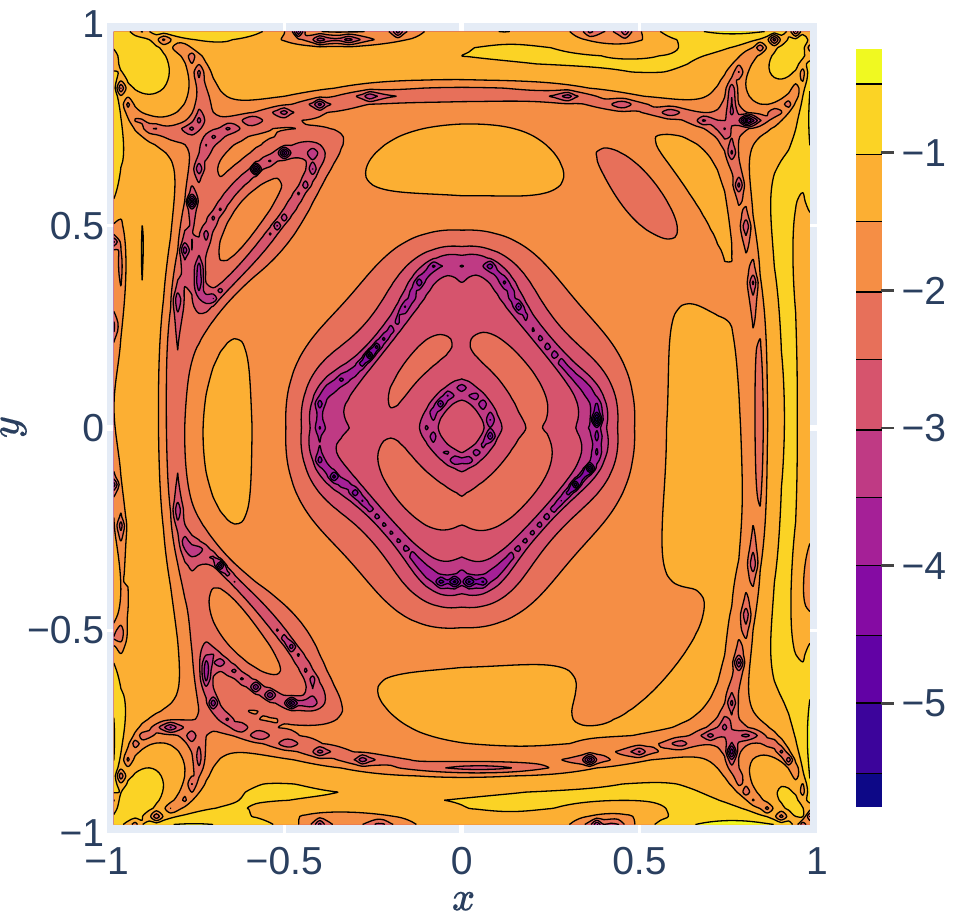}
         \caption{$\log_{10}(\epsilon_{\text{re}})$ - new }
         \label{fig:r_err_alpha_0.5_mod_loss_new}
     \end{subfigure}
    \caption{The contour plots of the absolute and relative errors of the DL solutions computed by the original and the new trapz-PiNN, for FPE with O-U potential in 2D \crefrange{eq:OU}{eq:OU bdry} sampled at $t = 0.2$ for $\alpha = 0.5$. }
    \label{fig:mod loss}
\end{figure}

Let $U(x,t)$ denote the physical observation or the high-fidelity simulated data at $(x,t)\in\Omega_T$ and $\mathcal{B}$ be a training batch. The new loss function is defined by \begin{align} \label{eq:mod loss}
    \mathbf{L}_{\mathcal{B}}^{\text{new}}(\Theta) &\coloneqq  \frac{\lambda_1}{|\mathcal{B}|}\sum_{(x_j,t_k)\in\mathcal{B}}|\partial_t\hat{u}(x_j,t_k;\Theta) - \mathcal{L}_h\hat{u}(x_j,t_k;\Theta)|^2 \nonumber \\
    & + \frac{\lambda_2}{|\mathcal{B}|}\sum_{(x_j,t_k)\in\mathcal{B}}\left(\frac{|\hat{u}(x_j,t_k;\Theta) - U(x_j,t_k)| + \delta}{|U(x_j,t_k)|+\delta}\right)^2, 
\end{align} where $\lambda_1,\lambda_2>0$ are constants, $\delta = 10^{-6}$ if $U$ vanishes at some points in $\Omega_T$, otherwise $\delta = 0$. One can see that the new loss function $\mathbf{L}_{\mathcal{B}}^{\text{new}}$ is a weighted sum of the loss function $\mathbf{L}_{\mathcal{B}}$ in \cref{eq:loss_function} and the `mean-squared' pointwise relative errors. To avoid repetitive presentation of the results in the same nature, we only study the Fokker-Planck equation with O-U potential when $\alpha = 0.5$ in this subsection. We compare the DL solutions computed by trapz-PiNN with two loss functions through pointwise absolute and relative errors. To have a fair comparison, we remain using the same set of hyperparameters chosen in \cref{subsection:OU}. We refer trapz-PiNNs equipped with the loss functions $\mathbf{L}_{\mathcal{B}}$ at \cref{eq:loss_function} and $\mathbf{L}_{\mathcal{B}}^{\text{new}}$ at \cref{eq:mod loss} as the original and the new trapz-PiNN, respectively. The high-fidelity simulated data are the FDM solutions evaluated at $t = 0.01 + \sfrac{k}{100},\ k = 0, \cdots, 19$. \cref{fig:mod loss} shows the comparison between the DL solutions predicted at $t = 0.2$ by the original and new trapz-PiNNs.

From \cref{fig:abs_err_alpha_0.5_mod_loss_original,,fig:abs_err_alpha_0.5_mod_loss_new}, we can see that the new trapz-PiNN reduces the maximum absolute error $\eps_{\text{abs}}$ from $0.01$ to $0.006$. Moreover, as shown in \cref{fig:r_err_alpha_0.5_mod_loss_original,,fig:r_err_alpha_0.5_mod_loss_new}, the pointwise relative errors $\eps_{\text{re}}$ for the new trapz-PiNN decreases to $3\%$ at majority of the total area  from the order of $10\%$ corresponding to the original trapz-PiNN. More precisely, $61.3\%$ of the total area has the pointwise relative error $\eps_{\text{re}}$ below $3\%$. In comparison, only $40.7\%$ of the area achieves such accuracy for the original trapz-PiNN. The $\mathcal{L}^2$ relative error $\veps$ for the DL solution predicted by the new trapz-PiNN is $5.8\times 10^{-3}$ while the original trapz-PiNN attains $\veps = 9.9\times 10^{-3}$. The range of $\veps$-accuracy of the new trapz-PiNN extended from $t = 0.2$ to at least beyond $t = 0.21$, as the $\mathcal{L}^2$ relative error at $t = 0.21$ is $\veps = 6.1\times 10^{-3}$.

The new loss function improves the performance of the trapz-PiNN on pointwise absolute and relative errors and extends the range of $\veps$-accuracy. Smaller maximum relative error can be achieved if larger NN shape and longer epoch are chosen. If physical observations or high-fidelity simulation is available, using the new loss function can achieve higher global and local accuracies. 

%%%%%%%%%%%%%%%%%%%%%%%%%%%%%%%%%%%%%%%%%%%%%%%%%%%%%%%%%%%%%%%%%%%%%%%%%%%%%%%%%%%% Mod loss plots

%%%%%%%%%%%%%%%%%%%%%%%%%%%%%%%%%%%%%%%%%%%%%%%%%%%%%%%%%%%%%%%%%%%%%%%%%%%%%%%%%%%%%%%%%%%%%%%%%%%%%%%%%%%%%%%%%%%%%%%%%%%%%%%%%%%%%%%%%%%%%%%%%%%%%%%%%%%%%%%%%%%%%%%%%%%%%%%%%%%%%%%%%%%%%%%%%%%%%%%%%%%%%%%%%%%%%%%%%%%%%%%%%%%%%%%%%%%%%%%%%%%%%%%%%%%%%%%%%%%%%%%%%%%%%%%%%%%%%%%%%%%%%%%%%%%%%%%%%%%%%%%%%%%%%%%%%%%%
\section{Conclusion}\label{sec:conclution}
We propose trapz-PiNN, a new physics-informed neural network, based on a recently developed modified trapezoidal rule, to solve non-local Fokker-Planck equations involving fractional laplacian. We have presented the simplest version of the modified trapezoidal rule in $\R^n$ and have verified second order accuracy for computing fractional laplacian in 2D. We have demonstrated trapz-PiNNs have high expressive power through numerical examples on fractional heat equations in 2D and 3D and Fokker-Planck eqution with O-U potential in 2D. The DL solutions that has low $\mathcal{L}^2$ relative error ($\veps$-accuracy) in general garantuee a small pointwise relative error for areas with high or moderate magnitude in $\Omega$. We also observe that trapz-PiNNs have some range of $\veps$-accuracy for almost all cases we studied. If physical oberservation or high-fidelity simulation of true solution is available, we propose an effective loss function integrating the extra information so that trapz-PiNN improves the performance on local metrics such as pointwise absolute and relative errors.

There are numerous questions remain to be investigated in future. Without resorting to using more brute force training or adopting external information, we will focus on designing an effective loss function to control the relative errors from region with vanishing magnitude, especially for PDEs with multi-scale solution profile. Due to RAM restriction, we only use moderate space resolution in solving fractional heat equation in 3D at \cref{subsubsection:heat 3D}. Low-memory fast algorithm for numerical fractional laplacian by modified trapezoidal rule has been developed and implemented in \cite{doi:10.1137/12086491X, HansenHa}. One can incorporate this fast algorithm into deep learning algorithm to alleviate the ``curse of dimensionality". We are also interested in developing efficient trapz-PiNNs for Fokker-Planck equation \cref{eq:Fokker-Planck} with discontinuous initial condition at boundary or with natural boundary (unbounded) condition.

\section{Acknowledgement}
The authors thanks Yiwei Wang for very helpful discussion about this work.

\bibliographystyle{plain}
\bibliography{main}

\end{document}